\pdfoutput=1
\documentclass[aps,pra,onecolumn,superscriptaddress,nofootinbib,floatfix]{revtex4-2}

\ifdefined\pdfgentounicode
  \input{glyphtounicode}
\fi

\usepackage{graphicx}
\usepackage{amsmath,amssymb}
\usepackage{xcolor}
\usepackage{tikz}
\usepackage[colorlinks=true,linkcolor=blue,citecolor=blue,urlcolor=blue]{hyperref}

\graphicspath{{figures/}}

\renewcommand{\thesection}{\arabic{section}}
\makeatletter
\def\p@subsection{\thesection.}
\makeatother

\definecolor{qubitgreen}{HTML}{00B050}
\definecolor{lightgreen}{HTML}{92D050}
\definecolor{kerrorange}{HTML}{F79646}
\definecolor{lightblue}{HTML}{00B0F0}

\newcommand{\eprfill}[3]{\fill[#3] ({3.7+3.2*(#1-1)},#2) rectangle ({6.9+3.2*(#1-1)},#2+1);}
\newcommand{\eprtext}[3]{\node at ({5.3+3.2*(#1-1)},#2+0.5) {#3};}
\newcommand{\eprlabel}[2]{\node[anchor=west] at (0.12,#1+0.5) {\textbf{#2}};}
\newcommand{\eprgrid}{%
  \draw (0,0) rectangle (13.3,5);
  \foreach \y in {1,...,4} \draw (0,\y) -- (13.3,\y);
  \foreach \x in {3.7,6.9,10.1} \draw (\x,0) -- (\x,5);}

\begin{document}

\title{Design of Qubit Readout Circuit for Purcell-Rate Suppression by Two-Path Interference}

\author{Ahmad Salmanogli}
\thanks{Corresponding author.}
\affiliation{Ankara Yildirim Beyazit University, Engineering Faculty, Electrical and Electronic Department, Ankara, Turkey}
\affiliation{Iranian Quantum Technologies Research Center (IQTEC), Tehran, Iran}

\author{Hesam Zandi}
\thanks{Corresponding author.}
\affiliation{Faculty of Electrical Engineering, K.N. Toosi University of Technology, Tehran, Iran}
\affiliation{Iranian Quantum Technologies Research Center (IQTEC), Tehran, Iran}

\author{Mohsen Akbari}
\affiliation{Iranian Quantum Technologies Research Center (IQTEC), Tehran, Iran}
\affiliation{Quantum Optics Laboratory, Department of Physics, Kharazmi University, Tehran, Iran}

\begin{abstract}
Dispersive readout of a superconducting qubit is limited by Purcell decay through the measurement port. We analyze a three-mode chain: a transmon coupled to a Purcell-filter resonator, which is coupled to a ported readout resonator. The layout studied here provides a direct capacitive path from the qubit to the ported mode; retaining it gives the closed-form rate $\Gamma_{\mathrm{P}} = \kappa_r(gg_k + g_q\Delta_{qf})^2/|\Delta_{qf}(\Delta_{qr} + i\kappa_r/2) - g_k^2|^2$, which vanishes identically when $gg_k + g_q\Delta_{qf} = 0$. This condition represents an exact two-path interference mechanism. It reproduces the bandpass-filter result reported by Sete et al.\ in the absence of the direct path and recovers the bare Purcell decay rate when the inter-mode coupling is removed. The resulting rate remains finite in the vacuum limit and vanishes when the qubit is fully decoupled. These limiting behaviors are independently confirmed by four numerical calculations and by the junction-admittance analysis. An energy-participation-ratio analysis yields a direct qubit--readout coupling of 5.05~MHz, with a corresponding cross-Kerr rate of 0.006~MHz. This coupling is 65\% of the 7.74~MHz value required to satisfy the cancellation condition within the rotating-wave approximation and 95\% of the 5.34~MHz required by the exact lumped-circuit solution. The dispersive cavity pull obtained from a Schrieffer--Wolff transformation that retains the transmon anharmonicity agrees with the corresponding exact-diagonalization result to within 5\%. The same model is evaluated for two layouts under matched readout constraints: the conventional configuration (Sys~I) and the proposed configuration (Sys~II). The two systems differ in both topology and frequency allocation, with the internal modes separated by 770~MHz in Sys~I and 220~MHz in Sys~II. The stronger hybridization in Sys~II produces a cavity pull of $2\chi_r/2\pi = -0.698$~MHz for the ported mode, compared with ($-0.076$)~MHz for Sys~I, while reducing the assignment error from $3.6\times10^{-3}$ against $9.1\times10^{-3}$. Imposing the interference condition on this layout suppresses the radiative decay channel, leaving $T_1$ limited by the intrinsic relaxation time of 60~$\mu$s and the filter's internal loss, which corresponds to 57.5~$\mu$s for a filter quality factor of $3.8\times10^{5}$. Under these conditions, the single-shot assignment error is $3.5\times10^{-3}$ for a 280~ns readout duration. Gaussian-state analysis further shows that the two internal modes remain separable in the steady-state operating regime of both layouts. Entanglement arises only as a transient counter-rotating effect; consequently, these correlations reflect the degree of mode hybridization but do not contribute to the readout error budget.

\end{abstract}

\maketitle

\section{Introduction}
\label{sec:intro}

Dispersive measurement is the standard readout of circuit quantum electrodynamics: a qubit is coupled off-resonantly to a microwave resonator, the resonator frequency is pulled by an amount that depends on the qubit state, and that pull is read out through a transmission line~\cite{r1,r3,r5,r26}. The same coupling that carries the signal out also carries the qubit excitation out. The resulting radiative relaxation, the Purcell effect~\cite{r2}, sets a ceiling on the qubit lifetime that scales as the square of the qubit--resonator coupling divided by the detuning, so that the two obvious remedies---reducing the coupling or increasing the detuning---both slow the measurement down~\cite{r4,r6,r7}. Resolving this trade-off without paying for it in measurement time is the purpose of a Purcell filter. The trade-off matters beyond a single device, because fast high-fidelity measurement is what limits multi-qubit algorithms and error correction in scalable processors~\cite{r9,r10,r11,r27,r28}. A Purcell filter is a frequency-selective network placed between the qubit--resonator system and the measurement line. It presents a large impedance mismatch at the qubit frequency, which suppresses the decay channel, and remains transparent at the readout frequency, which preserves the signal~\cite{r12,r13}. The bandpass realisation and its quantum theory were given by Sete, Martinis and Korotkov~\cite{r7}, following the tunable-coupling and fast-reset variants of Refs.~\cite{r12,r14}, and a series of experiments has since demonstrated combined readout/filter circuits~\cite{r15}, intrinsic filters built from the higher modes of the readout resonator itself~\cite{r30}, nonlinear filters that deactivate during the readout pulse~\cite{r32}, and multiplexed intrinsic-filter readout with measured time and error budgets~\cite{r33}. Two of these works are directly relevant here. Yen et al.~\cite{r34} suppress resonator-mediated decay by coupling the readout resonator to the feedline at two points, so that the two emission amplitudes interfere destructively, and Spring et al.~\cite{r33} obtain a notch of the same kind from combined capacitive and inductive coupling. Destructive interference between two emission paths is therefore an established route to Purcell protection, and it is the mechanism that the present work makes quantitative for a general three-mode chain.

This paper analyzes the circuit shown in Fig.~\ref{fig:layout}, in which a transmon qubit is coupled to a filter resonator, which is subsequently coupled to a driven and damped readout resonator. Section~\ref{sec:model} quantizes the circuit and formulates the master equation and quantum Langevin equations, with all relevant rates defined consistently and introduced. Section~\ref{sec:purcell} derives the radiative decay rate from the qubit self-energy, establishes that, for a purely resonator-mediated chain, exchanging the two internal modes is physically equivalent to relabeling them, and shows that the direct qubit-to-port coupling path enabled by the proposed layout can suppress the radiative decay rate to zero along a one-parameter manifold. Section~\ref{sec:dispersive} derives the dispersive frequency shift using a Schrieffer--Wolff transformation~\cite{r43,r44} while explicitly retaining the transmon anharmonicity. It distinguishes the cavity pull from the ac-Stark shift and clarifies the physical interpretation of the critical photon number. Section~\ref{sec:em} describes the electromagnetic design and specifies the circuit and simulation parameters required for reproducibility. Section~\ref{sec:readout} presents the readout simulations and a complete single-shot error budget under consistently matched experimental constraints. Section~\ref{sec:corr} examines the correlations between the two internal modes. Finally, the paper discusses the capabilities and limitations of the proposed design. Four appendices provide the detailed derivations, numerical cross-checks, and convergence studies supporting the main results. The scope of the analysis should be stated at the outset. The readout considered here is a linear dispersive measurement with a coherent drive. Within the rotating-wave Hamiltonian of Eq.~\ref{eq:Hrwa} the resonator part is quadratic and excitation-number conserving (a beam-splitter coupling $g_k(a^+b + ab^+)$ plus a linear drive), so that, conditioned on the qubit state, both resonator fields remain coherent states. The pointer states therefore carry no quadrature squeezing and no Wigner negativity, and their distinguishability is set entirely by the separation of two coherent states, $|\alpha_e(t) - \alpha_g(t)|$, through Eq.~\ref{eq:snr}. Readout schemes that use nonclassical light to go below this coherent-state limit, such as squeezed light injected into the readout cavity~\cite{r49} or two-mode squeezing generated and recombined by Josephson parametric converters~\cite{r31}, require an additional parametric resource that is not part of the circuit studied here. Such a resource is available in the measurement chain, since a Josephson parametric amplifier operated in phase-sensitive mode produces squeezed vacuum~\cite{r48}; combining it with the present circuit is left for future work. Accordingly, no nonclassicality-based readout advantage is claimed in this paper, and the inter-mode correlations of Section~\ref{sec:corr} are reported only as a diagnostic of the mode hybridization.

\section{Theoretical background}
\label{sec:theory}

\subsection{Circuit quantisation and the model Hamiltonian}
\label{sec:model}

\begin{figure}[tbp]
\centering
\includegraphics[width=0.68\linewidth]{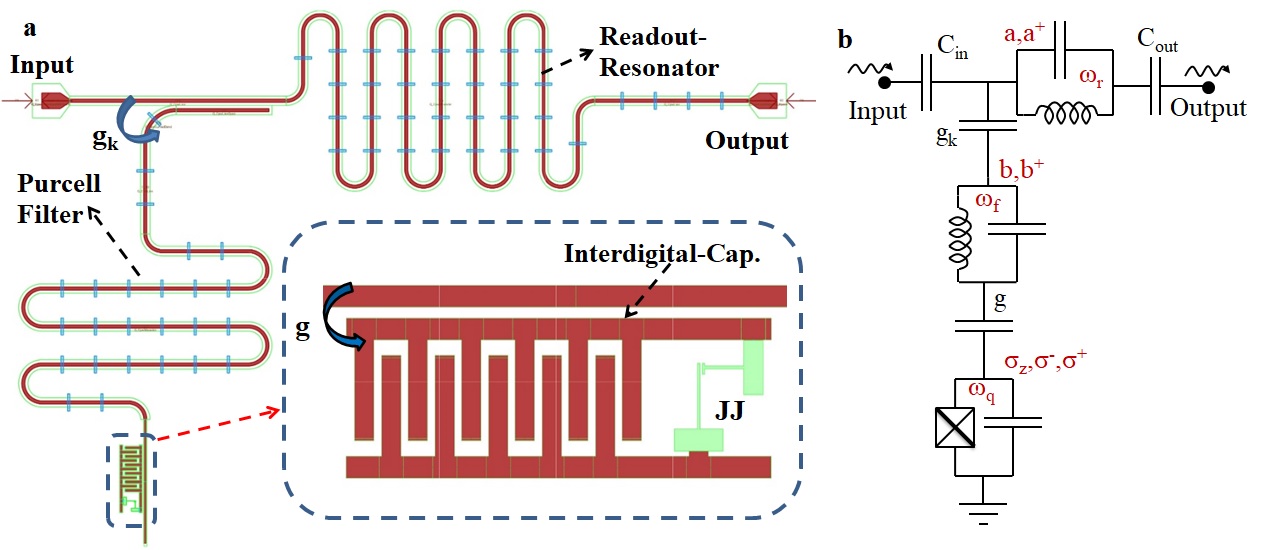}
\caption{a) Layout of the circuit, containing the readout resonator, the Purcell filter, the interdigital coupling capacitor and the Josephson junction, magnified in the inset; b) the lumped-element equivalent circuit. The transmon island has a finite capacitance to the readout resonator as well as to the filter, so the network is not a pure chain; the resulting direct coupling $g_q$ is retained throughout and is the origin of the interference condition.}
\label{fig:layout}
\end{figure}

The circuit is shown in Fig.~\ref{fig:layout}a and its lumped-element equivalent in Fig.~\ref{fig:layout}b. A transmon qubit with frequency $\omega_q$ and anharmonicity $\alpha$ is coupled capacitively to a filter resonator of frequency $\omega_f$, the filter is coupled to a readout resonator of frequency $\omega_r$, and the readout resonator is coupled to the measurement line.

In the following, two features of the layout are particularly important for the analysis. First, the coupling-capacitance network does not form a simple chain. The transmon island has a finite capacitive coupling to the readout resonator in addition to its coupling to the filter resonator. Consequently, a direct qubit-to-readout coupling, ($g_q$), exists alongside the indirect coupling mediated through the filter with strength ($g_k$). This direct coupling is not merely a parasitic effect that can be neglected; as demonstrated in Section~\ref{sec:purcell}, it is the key feature responsible for the difference between the proposed design and a conventional filter-coupled architecture. Second, the measurement port is coupled directly to the readout resonator, making this mode the only mode subject to the external measurement-induced linewidth ($\kappa_r$). The external decay rate therefore belongs exclusively to the readout mode and should not be assigned to the transmon or filter mode. Modeling the transmon as a weakly anharmonic (Duffing) oscillator~\cite{r6,r39} with annihilation operator ($q$), and denoting the readout and filter modes by ($a$) and ($b$), respectively, the quantized circuit Hamiltonian can be written as~\cite{r1,r26} ($\hbar = 1$ throughout the article):
\begin{equation}
\begin{split}
H ={}& \omega_q q^+q + \frac{\alpha}{2}q^+q^+qq + \omega_f b^+b + \omega_r a^+a\\
&+ g(q^+ + q)(b^+ + b) + g_q(q^+ + q)(a^+ + a) + g_k(a^+ + a)(b^+ + b)
\end{split}
\label{eq:H}
\end{equation}
where $g$, $g_q$ and $g_k$ are the qubit--filter, qubit--readout and filter--readout coupling rates. Eq.~\ref{eq:H} keeps the full position--position form of every coupling; no rotating-wave approximation has yet been made. Because all three couplings are small compared with the mode frequencies ($g/\omega_q \approx 0.02$ at the design point of Table~\ref{tab:params}), the counter-rotating terms may be dropped for the steady-state quantities computed in Sections~\ref{sec:purcell}, \ref{sec:dispersive} and \ref{sec:em}. Moving to a frame rotating at the readout drive frequency $\omega_d$ and applying the rotating-wave approximation~\cite{r16,r26}, the equation becomes:
\begin{equation}
\begin{split}
H_{\mathrm{RWA}} ={}& \Delta_q q^+q + \frac{\alpha}{2}q^+q^+qq + \Delta_f b^+b + \Delta_r a^+a\\
&+ g(q^+b + qb^+) + g_q(q^+a + qa^+) + g_k(a^+b + ab^+) + \varepsilon(a^+ + a)
\end{split}
\label{eq:Hrwa}
\end{equation}
with the drive-frame detunings $\Delta_q = \omega_q - \omega_d$, $\Delta_f = \omega_f - \omega_d$ and $\Delta_r = \omega_r - \omega_d$, and the readout drive amplitude $\varepsilon$. The qubit--mode detunings that control the radiative decay and the dispersive pull are separate quantities and are written $\Delta_{qf} = \omega_q - \omega_f$ and $\Delta_{qr} = \omega_q - \omega_r$. Section~\ref{sec:corr} reports results obtained both with and without the rotating-wave approximation, because the transient correlations discussed there are produced entirely by the counter-rotating terms of Eq.~\ref{eq:H}. Dissipation is introduced once, through Lindblad dissipators~\cite{r40,r41}, and the environment is not represented by any additional bosonic mode. The master equation is introduced as:
\begin{equation}
\dot{\rho} = -i[H_{\mathrm{RWA}},\rho] + \kappa_r(1 + n_{\mathrm{th}})\mathcal{D}[a]\rho + \kappa_r n_{\mathrm{th}}\mathcal{D}[a^+]\rho + \kappa_f^{\mathrm{int}}\mathcal{D}[b]\rho + \gamma_{\mathrm{int}}\mathcal{D}[q]\rho + 2\gamma_\phi\mathcal{D}[q^+q]\rho
\label{eq:master}
\end{equation}
with $\mathcal{D}[c]\rho = c\rho c^+ - \tfrac{1}{2}\{c^+c,\rho\}$. Here, $\kappa_r$ denotes the external linewidth of the readout resonator, corresponding to its coupling to the measurement port. $\kappa_f^{\mathrm{int}}$ represents the internal (dielectric) loss rate of the filter mode, while $\gamma_{\mathrm{int}}$ denotes the intrinsic, non-radiative relaxation rate of the qubit. The parameter $\gamma_\phi$ represents the pure-dephasing rate, and $n_{\mathrm{th}}$ denotes the residual thermal occupation of the measurement port. Each rate is introduced once and retains the same definition and physical interpretation throughout the paper. The effective port-induced linewidth of the filter mode, which is a derived quantity, is denoted by $\kappa_f^{\mathrm{eff}}$ and is defined explicitly in Eq.~\ref{eq:sete}. The corresponding quantum Langevin equations, obtained from Eq.~\ref{eq:Hrwa} together with the standard input--output relations~\cite{r16,r42}, are
\begin{equation}
\left\{
\begin{aligned}
\dot{a} &= -i\Delta_r a - ig_k b - ig_q\sigma^- - \kappa_r a/2 - i\varepsilon - \sqrt{\kappa_r}\,a_{\mathrm{in}}\\
\dot{b} &= -i\Delta_f b - ig_k a - ig\sigma^- - \kappa_f^{\mathrm{int}}b/2 - \sqrt{\kappa_f^{\mathrm{int}}}\,b_{\mathrm{in}}\\
\dot{\sigma}^- &= -i\Delta_q\sigma^- + ig\sigma_z b + ig_q\sigma_z a - \gamma_2\sigma^-/2
\end{aligned}
\right.
\label{eq:langevin}
\end{equation}
where $\gamma_2 = \gamma_{\mathrm{int}}/2 + \gamma_\phi + \Gamma_{\mathrm{P}}$ is the total transverse decay rate, $a_{\mathrm{in}}$ and $b_{\mathrm{in}}$ are the vacuum input operators of the port and of the filter loss channel, and $\sigma^-$, $\sigma_z$ are the two-level reduction of $q$ used when the transmon is restricted to its lowest two levels. The filter-mode damping, qubit transverse decay, and input-noise operators are retained explicitly throughout the formulation, ensuring that Eq.~\ref{eq:langevin} represents the same open quantum system described by the master equation in Eq.~\ref{eq:master}. The corresponding output field is given by $a_{\mathrm{out}} = a_{\mathrm{in}} + \sqrt{\kappa_r}\,a$, which is the observable integrated in Section~\ref{sec:readout}. Table~\ref{tab:params} provides a complete definition of all symbols, including their physical meanings and design-point values, thereby making the model self-contained and unambiguous without requiring additional information from the surrounding text.

\subsection{Radiative decay rate and the interference condition}
\label{sec:purcell}

\begin{figure}[tbp]
\centering
\includegraphics[width=0.88\linewidth]{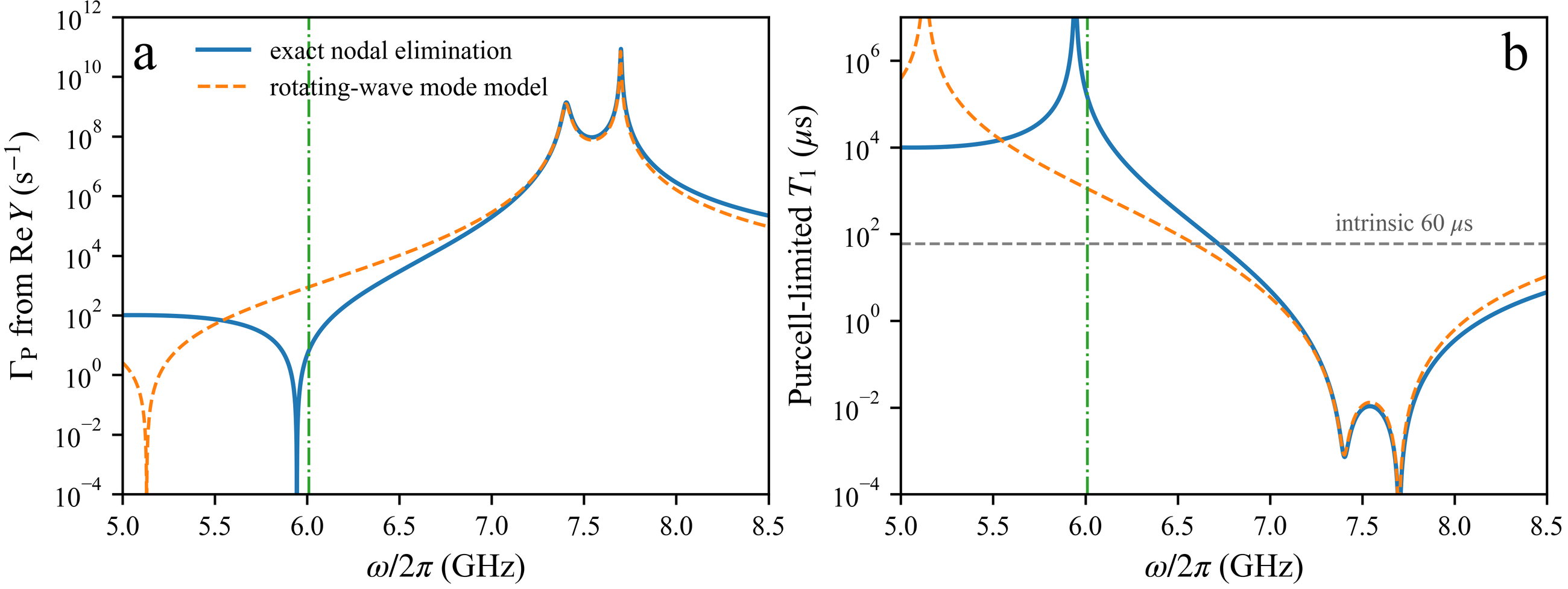}
\caption{(a) Purcell rate of the proposed circuit (Sys~II) at the design point of Table~\ref{tab:params}, obtained from the real part of the admittance seen by the Josephson junction by exact nodal elimination of the network of Fig.~\ref{fig:layout}b (solid), compared with the rotating-wave mode model of Eq.~\ref{eq:Hrwa} (dashed); (b) the corresponding Purcell-limited $T_1$. The vertical line marks the qubit frequency and the horizontal line the intrinsic 60~$\mu$s. The notch lies at 5.94~GHz in the exact circuit and at 5.13~GHz in the rotating-wave model; the difference is the factor $F$ discussed below Eq.~\ref{eq:admnull}. Only the port contribution is shown; the internal-loss floor of the filter is discussed in Section~\ref{sec:purcell} and shown in Fig.~\ref{fig:A1}.}
\label{fig:admittance}
\end{figure}

The radiative decay rate of the qubit is determined from the self-energy induced by the coupled two-mode network~\cite{r4,r7,r26}. The qubit--mode coupling vector be defined as $\mathbf{g} = (g, g_q)^{\mathrm{T}}$, and $H_c$ denote the corresponding non-Hermitian mode matrix, which incorporates the port-induced linewidth of the readout mode. Under this formulation, the self-energy provides the effective dissipative contribution of the coupled network to the qubit dynamics introduced as:
\begin{equation}
\Sigma(\omega) = \mathbf{g}^{\mathrm{T}}(\omega\mathbf{1} - H_c)^{-1}\mathbf{g},\qquad
\Gamma_{\mathrm{P}} = -2\,\operatorname{Im}\,\Sigma(\omega_q),\qquad
H_c = \begin{pmatrix} \omega_f - i\kappa_f^{\mathrm{int}}/2 & g_k\\ g_k & \omega_r - i\kappa_r/2 \end{pmatrix}
\label{eq:selfenergy}
\end{equation}
Inverting the $2\times2$ matrix (Appendix~\ref{app:A}) and taking the imaginary part gives the central result of this paper written for $\kappa_f^{\mathrm{int}} \ll \kappa_r$ as:
\begin{equation}
\Gamma_{\mathrm{P}} = \kappa_r(gg_k + g_q\Delta_{qf})^2\big/\left|\Delta_{qf}(\Delta_{qr} + i\kappa_r/2) - g_k^2\right|^2,\qquad
\Delta_{qf} = \omega_q - \omega_f,\quad \Delta_{qr} = \omega_q - \omega_r
\label{eq:gammaP}
\end{equation}
The more general expression, which incorporates the internal loss of the filter mode within the same framework, is given in Eq.~\ref{eq:A4}. Eq.~\ref{eq:gammaP} is exact to second order in the two qubit--mode coupling strengths and to all orders in $g_k$ and $\kappa_r$. It also exhibits the expected properties of a physical decay rate. In particular, it vanishes for $g = g_q = 0$, corresponding to complete decoupling of the qubit from the circuit. Because the expression contains no explicit photon-number dependence, it remains finite in the vacuum limit and therefore describes the qubit relaxation rate when the drive is switched off. Moreover, it is proportional to the dissipative rate $\kappa_r$ of the mode coupled to the measurement port, as required for a radiative decay process. The resulting rate depends on both qubit--mode coupling strengths and on the detunings of the qubit from the two relevant modes. Two limits identify the expression: First setting $g_q = 0$, so that the circuit is a pure chain; Second assuming $g_k^2 \ll |\Delta_{qf}\Delta_{qr}|$ gives:
\begin{equation}
\Gamma_{\mathrm{P}} \simeq (g/\Delta_{qf})^2\,\kappa_f^{\mathrm{eff}}(\omega_q),\qquad
\kappa_f^{\mathrm{eff}}(\omega) = g_k^2\kappa_r\big/\left[(\omega - \omega_r)^2 + \kappa_r^2/4\right]
\label{eq:sete}
\end{equation}
This expression corresponds to the bandpass-filter result reported by Sete et al.\ (Eq.~31 of Ref.~\cite{r7}), in which the qubit inherits the linewidth induced by the measurement port on the mode to which it is coupled, with a weighting factor of $(g/\Delta_{qf})^2$. In the limiting case $g_k = 0$, the expression reduces to $\Gamma_{\mathrm{P}} = g_q^2\kappa_r/(\Delta_{qr}^2 + \kappa_r^2/4)$, which is the conventional bare Purcell decay rate for a qubit directly coupled to a damped resonator~\cite{r4}. The numerator of Eq.~\ref{eq:gammaP} represents a genuine physical interference effect rather than a simple relabelling of parameters. The qubit couples to the measurement port through two coherent pathways: (i) an indirect pathway mediated by the filter, followed by the coupling $g_k$, which contributes a term proportional to $g\,g_k$, and (ii) a direct qubit--port pathway, which contributes a term proportional to $g_q\Delta_{qf}$. These add coherently, and because $\Delta_{qf}$ is negative for a qubit below the filter, the two contributions have opposite signs for positive couplings. The radiative rate therefore vanishes identically on the surface as:
\begin{equation}
gg_k + g_q\Delta_{qf} = 0 \quad\Leftrightarrow\quad g_q^{\mathrm{null}} = -gg_k/\Delta_{qf} \quad\Leftrightarrow\quad \omega_{\mathrm{notch}} = \omega_f - gg_k/g_q
\label{eq:null}
\end{equation}
The first term is a condition on the couplings at fixed frequencies; the second gives the direct coupling that a given chain requires; the third gives the qubit frequency at which a given set of couplings produces a notch. This is the same mechanism that Yen et al.~\cite{r34} realize by coupling the readout resonator to the feedline at two points and Spring et al.~\cite{r33} realize with combined capacitive and inductive coupling. The contribution here is the closed-form condition for a general three-mode chain with a direct qubit-to-port path, its equivalence to a null of the circuit admittance (discussed in the following), and the observation that the layout of Fig.~\ref{fig:layout} already supplies a large fraction of the required direct coupling. The last point is quantitative. The energy-participation-ratio (EPR) analysis of Sys~II in Table~\ref{tab:epr}b gives a qubit--readout-resonator cross-Kerr of 0.006~MHz. Inverting the transmon relation $\chi = 2g^2\alpha/[\Delta(\Delta+\alpha)]$ gives $g_q/2\pi = 5.05$~MHz, while the qubit--filter cross-Kerr of 2.95~MHz gives $g/2\pi = 127.8$~MHz. At the design point of Table~\ref{tab:params} the null of Eq.~\ref{eq:null} requires $g_q/2\pi = 7.74$~MHz, so the layout is already at 65\% of the rotating-wave cancellation condition; the exact solution of the lumped circuit discussed below Eq.~\ref{eq:admnull} lowers the requirement to 5.34~MHz, of which the layout already supplies 95\%. In terms of the quantity the electromagnetic solver reports, the qubit--readout cross-Kerr would have to be raised from 0.006~MHz to 0.014~MHz in the rotating-wave approximation, or to 0.0067~MHz with the full-circuit correction. Either is a modest change of one coupling capacitance, and it is the concrete design target that follows from this analysis. Fig.~\ref{fig:A1} shows $\Gamma_{\mathrm{P}}$ as a function of $g_q$ computed by five independent routes: Eq.~\ref{eq:gammaP}; an equivalent port-amplitude form, in which the qubit-to-port transition amplitude is obtained by solving the two-mode linear system and $\Gamma_{\mathrm{P}}$ is $\kappa_r$ times its squared modulus; the imaginary part of the complex eigenvalue of the non-Hermitian single-excitation matrix, which makes no expansion in the couplings at all; the same closed form evaluated at the Lamb-shifted qubit frequency; and a direct Lindblad simulation of the decay of $|e,0,0\rangle$. Away from the notch, Eq.~\ref{eq:A4} agrees with the exact eigenvalue to within 4\% for all circuits of Table~\ref{tab:circuits}, and Eq.~\ref{eq:gammaP} agrees to within 6\% with the exact eigenvalue evaluated at $\kappa_f^{\mathrm{int}} = 0$; the difference between Eq.~\ref{eq:gammaP} and the exact rate at finite $\kappa_f^{\mathrm{int}}$ is the filter-loss channel, which Eq.~\ref{eq:gammaP} omits by construction. At the notch the closed form gives zero by construction while the exact eigenvalue saturates at $7.4\times10^{2}$~s$^{-1}$, and this residue is not a failure of the cancellation: it equals $(g/\Delta_{qf})^2\kappa_f^{\mathrm{int}} = 7.5\times10^{2}$~s$^{-1}$, the dielectric loss of the filter mode inherited through the same dispersive weight. Setting $\kappa_f^{\mathrm{int}} = 0$ in the exact calculation returns 0.26~s$^{-1}$, so the interference null of Eq.~\ref{eq:null} survives to all orders in the couplings, and what limits it in practice is the internal quality factor of the filter, not the approximation used to derive it. The result also can be obtained without quantizing anything. The Purcell rate of a junction shunted by an admittance $Y(\omega)$ is $\Gamma_{\mathrm{P}} = \operatorname{Re}Y(\omega_q)/C_q$~\cite{r4,r38}, and eliminating the two resonator nodes of Fig.~\ref{fig:layout}b by a Schur complement (Appendix~\ref{app:B}) gives a real part that vanishes by considering:
\begin{equation}
\omega_q C_{c1}C_{fr} + C_{c2}\,y_f(\omega_q) = 0,\qquad
y_f(\omega) = \omega C_{f\Sigma}\left(1 - \omega_f^2/\omega^2\right)
\label{eq:admnull}
\end{equation}
where $C_{c1}$ and $C_{c2}$ are the qubit-to-filter and qubit-to-readout coupling capacitances, $C_{fr}$ couples the two resonators and $C_{f\Sigma}$ is the total capacitance of the filter node. The condition in Eq.~\ref{eq:admnull} represents the classical counterpart of Eq.~\ref{eq:null}, indicating that the two displacement currents injected by the junction into the port cancel each other. A comparison of the two expressions shows that the rotating-wave condition is modified by the factor $F = 2\omega_q^2/[\omega_f(\omega_q + \omega_f)] = 2\omega_f/(\omega_q + \omega_f) - 2(\omega_f - \omega_q)/\omega_f$. The first term is the counter-rotating correction neglected in Eq.~\ref{eq:Hrwa}. The second arises because, in the capacitance network, the series connection of $C_{c1}$ and $C_{fr}$ adds an electrostatic qubit--readout coupling $2gg_k/\omega_f$ to the direct one; $g_q$ in Eqs.~\ref{eq:null} and \ref{eq:admnull} is therefore the contribution of $C_{c2}$ alone, which is what the EPR analysis of the drawn layout returns (Section~\ref{sec:em}). Accordingly, the exact condition is given by $g\,g_k F + g_q\Delta_{qf} = 0$. At the design point, $\omega_q/2\pi = 6.010$~GHz and $\omega_f/2\pi = 7.660$~GHz, giving $F = 0.690$. The full-circuit solution therefore reduces the coupling required for exact cancellation by approximately one-third compared with the rotating-wave approximation, from $g_q/2\pi = 7.74$~MHz to 5.34~MHz, and the EPR value of 5.05~MHz is 95\% of the latter. The values reported in the following sections, including circuit C of Table~\ref{tab:circuits}, correspond to the rotating-wave condition given by Eq.~\ref{eq:null} and therefore do not include this correction. The difference between the two values provides an estimate of the systematic uncertainty associated with the design target; a layout that is to reach the null must be tuned with the full-circuit condition, since in the full circuit a direct coupling of 7.74~MHz overshoots the null and radiates at $4.6\times10^{2}$~s$^{-1}$. Fig.~\ref{fig:admittance} shows the interference condition in the frequency domain for the proposed circuit (Sys~II): the Purcell rate derived from $\operatorname{Re}\{Y(\omega)\}$ by exact nodal elimination (solid) and, for comparison, the same rate from the rotating-wave mode model of Eq.~\ref{eq:Hrwa} (dashed), together with the corresponding Purcell-limited $T_1$. In Fig.~\ref{fig:admittance}a, both curves peak at the two normal-mode frequencies of the filter--readout network, 7.40 and 7.70~GHz, and collapse in a deep notch where the two decay paths cancel. In the rotating-wave model the notch lies at 5.13~GHz, 0.88~GHz below $\omega_q$, and the rate at the qubit frequency is $\Gamma_{\mathrm{P}} = 9.0\times10^{2}$~s$^{-1}$, as in Table~\ref{tab:circuits}. In the exact circuit the notch moves up to 5.94~GHz, only 66~MHz below $\omega_q$, and the radiative rate at the qubit frequency falls to 7~s$^{-1}$, about two orders of magnitude below the internal-loss floor of the filter; for circuit B the rotating-wave values used in the rest of the paper are therefore conservative. Fig.~\ref{fig:admittance}b shows that the Purcell-limited $T_1$ exceeds the intrinsic 60~$\mu$s for all qubit frequencies below 6.72~GHz in the exact circuit (6.59~GHz in the rotating-wave model).

\subsection{Dispersive pull, ac-Stark shift and the critical photon number}
\label{sec:dispersive}

\begin{figure}[tbp]
\centering
\includegraphics[width=\linewidth]{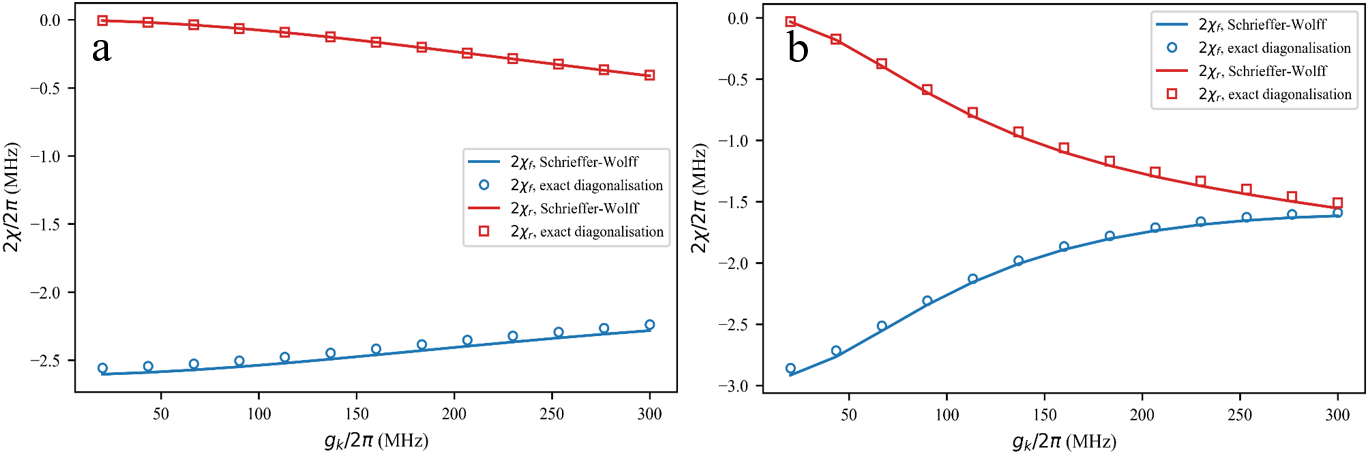}
\caption{Cavity pull of the filter mode (blue) and of the readout mode (red) as a function of the filter--readout coupling $g_k$ (a: Sys~I and b: Sys~II): lines, the Schrieffer--Wolff result of Eq.~\ref{eq:chi}; symbols, exact numerical diagonalization of the full transmon-plus-two-mode Hamiltonian. The percentages quoted are those of Sys~II, panel (b); for Sys~I, panel (a), the agreement is within 2.0\% and 1.1\%. At the design value $g_k/2\pi = 100$~MHz the ported mode carries 2.9\% of the total pull in Sys~I and 23\% of it in Sys~II.}
\label{fig:pull}
\end{figure}

The cavity pull and the ac-Stark shift are distinct quantities and are kept apart throughout this section. A dispersive pull must vanish when the qubit is decoupled from the circuit and when the transmon anharmonicity vanishes, and, being a property of the undriven spectrum, it cannot contain a drive-frame detuning. Because $g_k$ is not small compared with the filter--readout detuning, the two resonators must be diagonalized before the qubit is treated perturbatively. The linear $2\times2$ block of Eq.~\ref{eq:Hrwa} is brought to normal form by the rotation introduced as:
\begin{equation}
\begin{aligned}
&\tan 2\theta = 2g_k/(\omega_f - \omega_r),\qquad g_+ = g\cos\theta + g_q\sin\theta,\qquad g_- = -g\sin\theta + g_q\cos\theta\\
&\kappa_+ = \kappa_r\sin^2\theta,\qquad \kappa_- = \kappa_r\cos^2\theta
\end{aligned}
\label{eq:rotation}
\end{equation}
with normal-mode frequencies $\omega_\pm$. Eq.~\ref{eq:rotation} is the only place where the two resonators are treated non-perturbatively; everything after it is second order in the qubit couplings. A Schrieffer--Wolff transformation~\cite{r43,r44} applied to each normal mode, keeping the transmon as a multilevel Duffing oscillator~\cite{r6,r39,r45}, then gives the cavity pull by:
\begin{equation}
2\chi_\pm = 2g_\pm^2\alpha\big/\left[\Delta_\pm(\Delta_\pm + \alpha)\right],\qquad
\Delta_\pm = \omega_q - \omega_\pm,\qquad
2\chi_r = 2\chi_+\sin^2\theta + 2\chi_-\cos^2\theta
\label{eq:chi}
\end{equation}
The last expression projects the normal-mode pulls back onto the readout resonator, which is the mode the measurement port senses. Eq.~\ref{eq:chi} has the properties that a dispersive pull must possess: it vanishes when the qubit couplings vanish, it vanishes when the anharmonicity vanishes, since a harmonic qubit produces no state-dependent pull, and it reduces to the familiar $2g^2\alpha/[\Delta(\Delta+\alpha)]$ of a single mode when $\theta \to 0$. At the design point of Table~\ref{tab:params} it gives $2\chi_f/2\pi = -2.26$~MHz on the filter mode and $2\chi_r/2\pi = -0.70$~MHz on the readout resonator. Fig.~\ref{fig:pull} compares the analytical results with exact numerical diagonalization of the full transmon--two-mode Hamiltonian over the range $20 \le g_k/2\pi \le 300$~MHz. Figure~\ref{fig:pull}(a) presents the conventional configuration (Sys~I), while Fig.~\ref{fig:pull}(b) shows the corresponding results for the proposed configuration (Sys~II); the two circuit architectures are illustrated in Fig.~\ref{fig:layouts}. Equation~(\ref{eq:chi}) agrees closely with the exact diagonalization for both systems. For Sys~I, the deviations are within 2.0\% for the filter-mode dispersive pull and 1.1\% for the readout-mode pull, compared with 2.1\% and 4.9\%, respectively, for Sys~II. Thus, the 5\% accuracy claimed for Eq.~(\ref{eq:chi}) is maintained across both architectures. The key distinction between the two systems is not the total dispersive pull but how that pull is distributed between the two hybridized modes. In both configurations, the quantity $|2\chi_f| + |2\chi_r|$ remains nearly independent of $g_k$. As $g_k/2\pi$ increases from 20 to 300~MHz, it changes only from 2.61 to 2.70~MHz in Sys~I and from 2.95 to 3.17~MHz in Sys~II. Thus, increasing the filter--readout coupling does not generate additional dispersive pull; instead, it primarily redistributes the existing qubit-induced pull between the two normal modes.

This redistribution is comparatively weak in Sys~I, where the two internal modes are separated by 770~MHz. The readout-mode contribution $|2\chi_r|/2\pi$ increases from only 0.007~MHz at $g_k/2\pi = 20$~MHz to 0.41~MHz at 300~MHz. At the design value $g_k/2\pi = 100$~MHz, the port-coupled mode therefore carries only 2.9\% of the total dispersive pull. In Sys~II, where the internal modes are separated by only 220~MHz, the redistribution is substantially stronger: $|2\chi_r|/2\pi$ reaches 0.41~MHz already at $g_k/2\pi \sim 70$~MHz, while at the design point the ported mode carries approximately 23\% of the total pull. At $g_k/2\pi = 200$~MHz, the two contributions become nearly equal, with $2\chi_f/2\pi = -1.62$~MHz and $2\chi_r/2\pi = -1.56$~MHz. The approximately ninefold difference at the design point---$2\chi_r/2\pi = -0.076$~MHz for Sys~I versus $-0.698$~MHz for Sys~II, as reported in Table~\ref{tab:circuits}---therefore does not originate from the nominal assignment of one resonator as the filter and the other as the readout mode. Rather, it results from the frequency separation of the two internal modes. When the modes are sufficiently close, the rotation described by Eq.~\ref{eq:rotation} more efficiently transfers the qubit-induced dispersive pull to the normal mode coupled to the measurement port. This enhanced port-visible pull directly underlies the difference in single-shot assignment error reported in Table~\ref{tab:circuits}.

However, the ac-Stark shift is a different quantity: it is the shift of the qubit frequency produced by the photons already in the resonator~\cite{r8}, $\delta\omega_q = 2\chi_r\bar{n}_r$, to which is added the dispersive Lamb shift, which is present even in vacuum. At the design point the Lamb shift of the qubit is $-9.8$~MHz and the ac-Stark shift at $\bar{n}_r = 10$ photons is $-7.0$~MHz. Finally, the critical photon number~\cite{r1,r26,r46} is:
\begin{equation}
n_{\mathrm{crit}} = \min_{\pm}\left(\Delta_\pm^2/4g_\pm^2\right)
\label{eq:ncrit}
\end{equation}
It marks the photon number at which the dispersive expansion underlying Eq.~\ref{eq:chi} and Eq.~\ref{eq:ncrit} breaks down and qubit--resonator mixing becomes non-perturbative; it is not a measure of the nonclassicality of the circuit. All readout results below are quoted at $\bar{n} = 10$, i.e.\ at $\bar{n}/n_{\mathrm{crit}} \approx 0.2$ ($n_{\mathrm{crit}} \approx 49$ at the design point), where the dispersive treatment is valid. Because Eq.~\ref{eq:ncrit} is a minimum over both normal modes and the photon-number constraint of Section~\ref{sec:readout} is imposed on either mode, the dispersive limit is enforced on the filter-like mode as well as on the ported mode; in Sys~II it is in fact the filter-like normal mode ($\cos^2\theta = 0.87$) that sets $n_{\mathrm{crit}}$. Measurement-induced state transitions of the transmon can set in below $n_{\mathrm{crit}}$~\cite{r51}; they are not captured by the three-level truncation of Appendix~\ref{app:C} and would further limit the usable photon number.

\subsection{Electromagnetic design and reproducibility}
\label{sec:em}

Two coplanar-waveguide layouts were drawn and analyzed: Sys~I, in which the transmon is coupled to the readout resonator and the readout resonator to the ported Purcell filter, which is the conventional arrangement of Ref.~\cite{r7}; and Sys~II, the arrangement of Fig.~\ref{fig:layout}, in which the transmon is coupled to the filter and the filter to the ported readout resonator. Each layout was simulated with a finite-element electromagnetic solver and the eigenmodes were quantized by the EPR method~\cite{r37}, cross-checked against black-box quantisation~\cite{r38}; the analysis returns the mode frequencies, the transmon anharmonicity and the cross-Kerr matrix. Table~\ref{tab:epr} collects the results. They are two independent designs, each at its own operating point, so their frequencies and cross-Kerr values differ. However, the matched comparison of the two architectures is made in Section~\ref{sec:readout}, where the readout constraints---photon number, integration time, measurement efficiency and thermal population---are held identical by construction and the couplings are those extracted from Table~\ref{tab:epr}.

\begin{figure}[tbp]
\centering
\includegraphics[width=0.96\linewidth]{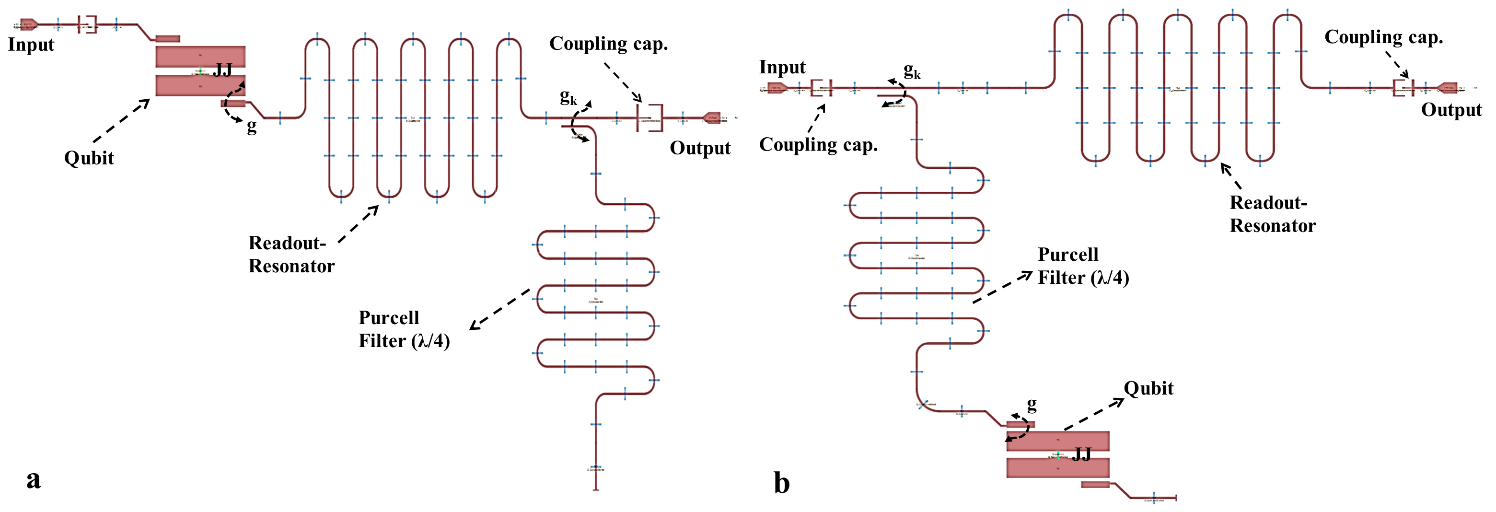}
\caption{Coplanar-waveguide layouts analysed in the electromagnetic solver: a) Sys~I, the conventional ordering, in which the transmon is coupled to the readout resonator and the readout resonator to the ported Purcell filter; b) Sys~II, the ordering of Fig.~\ref{fig:layout} as a proposed circuit in this work.}
\label{fig:layouts}
\end{figure}

\begin{figure}[tbp]
\centering
\includegraphics[width=0.53\linewidth]{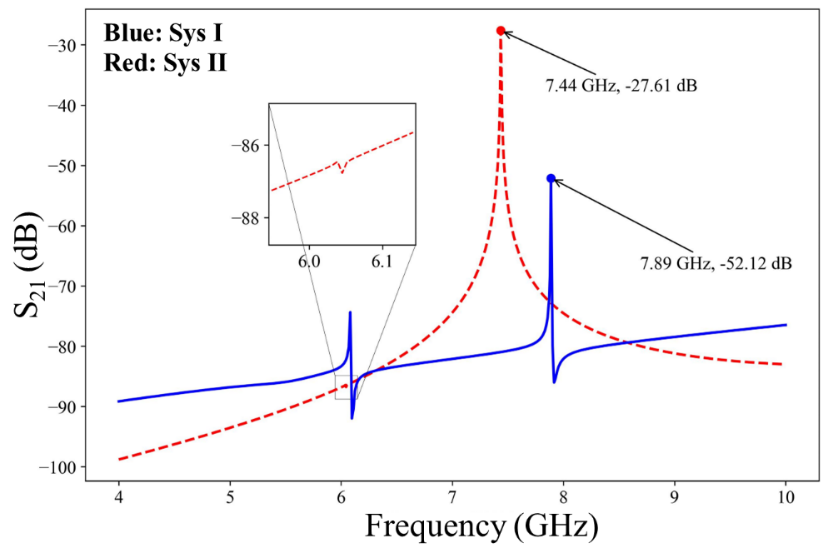}
\caption{Transmission spectrum ($S_{21}$) vs.\ frequency (GHz) for Sys~I (blue) and Sys~II (red). The suppression of the qubit resonance in Sys~II is a consistency check that the filter is opaque at the qubit frequency and transparent at the readout frequency. The spectra are those of the drawn layouts of Fig.~\ref{fig:layouts}, not of the design point of Table~\ref{tab:params}.}
\label{fig:s21}
\end{figure}

The cross-Kerr entries are the quantities from which the couplings used in Section~\ref{sec:theory} are obtained. Inverting $\chi_{ij} = 2g_{ij}^2\alpha/[\Delta_{ij}(\Delta_{ij} + \alpha)]$ for the qubit--mode entries of Table~\ref{tab:epr}b gives $g/2\pi = 127.8$~MHz for the qubit--filter coupling and $g_q/2\pi = 5.05$~MHz for the direct qubit--readout coupling; the corresponding numbers from Table~\ref{tab:epr}a are 133.6~MHz and 4.58~MHz. The small but non-zero qubit--readout entry in Sys~II, 0.006~MHz, is exactly the direct path of Section~\ref{sec:purcell}, and it is the reason the analysis of that section applies to this layout rather than to an idealized chain. One consequence of this extraction must be stated explicitly. The EPR analysis returns the eigenmodes of the drawn linear network, which are already hybridized by whatever filter--readout coupling the layout contains. Had the layout of Fig.~\ref{fig:layouts}b realized $g_k/2\pi = 100$~MHz, its two eigenmodes would be split by 297~MHz rather than the 220~MHz of Table~\ref{tab:epr}b, and the qubit pull on the readout-like eigenmode would be about 0.70~MHz rather than 0.006~MHz. The ratio of the two qubit--mode cross-Kerr entries, $0.006/2.95 \approx 0.002$, instead implies a drawn filter--readout coupling of order 10~MHz or less (hybridization alone would reproduce the 0.006~MHz entry at $g_k/2\pi \approx 7$~MHz), and the same holds for Sys~I in Table~\ref{tab:epr}a. Therefore (i) $g$ and $g_q$ are extracted from an almost unhybridized layout, so that the frequencies of Table~\ref{tab:epr} serve as the bare mode frequencies of the model; (ii) $g_k/2\pi = 100$~MHz is a design target that is added to the model and is not realized in the layouts of Fig.~\ref{fig:layouts}; and (iii) the transmission spectra of Fig.~\ref{fig:s21} refer to the drawn layouts, not to the design point of Table~\ref{tab:params}. Realizing $g_k$ requires enlarging the filter--readout coupling capacitance $C_{fr}$, which is how it enters the full-circuit condition of Section~\ref{sec:purcell}. It is notable to mention that all quantum results in this paper depend on the electromagnetic model only through five numbers---two mode frequencies, the anharmonicity and two qubit--mode cross-Kerr entries can reproduce every subsequent figure from Table~\ref{tab:params} alone.

\begin{table}[tbp]
\caption{Data resulting from the EPR analysis, presenting resonance frequencies, anharmonicities (qubit: dark green; Purcell filter and readout resonator: light blue) and cross-Kerr interactions (qubit--Purcell filter: light green; qubit--readout resonator: orange). The qubit--mode cross-Kerr entries are the quantities from which the couplings $g$ and $g_q$ of Table~\ref{tab:params} are obtained, by inverting $\chi = 2g^2\alpha/[\Delta(\Delta+\alpha)]$.}
\label{tab:epr}
\centering
a) Sys~I\\[4pt]
\begin{tikzpicture}[x=1cm,y=-0.62cm]
\eprfill{1}{2}{qubitgreen} \eprfill{2}{2}{lightgreen} \eprfill{3}{2}{kerrorange}
\eprfill{1}{3}{lightgreen} \eprfill{2}{3}{lightblue}
\eprfill{1}{4}{kerrorange} \eprfill{3}{4}{lightblue}
\eprgrid
\eprlabel{0}{Sys I} \eprtext{1}{0}{\textbf{Qubit}} \eprtext{2}{0}{\textbf{Purcell Filter Res.}} \eprtext{3}{0}{\textbf{Readout Res.}}
\eprlabel{1}{Frequency} \eprtext{1}{1}{6.02 GHz} \eprtext{2}{1}{8.66 GHz} \eprtext{3}{1}{7.89 GHz}
\eprlabel{2}{Qubit} \eprtext{1}{2}{296 MHz} \eprtext{2}{2}{0.0016 MHz} \eprtext{3}{2}{2.61 MHz}
\eprlabel{3}{Purcell Filter Res.} \eprtext{1}{3}{0.0016 MHz} \eprtext{2}{3}{1.6 kHz} \eprtext{3}{3}{1.86e-4 MHz}
\eprlabel{4}{Readout Res.} \eprtext{1}{4}{2.61 MHz} \eprtext{2}{4}{1.86e-4 MHz} \eprtext{3}{4}{5.35 kHz}
\end{tikzpicture}

\vspace{10pt}
b) Sys~II\\[4pt]
\begin{tikzpicture}[x=1cm,y=-0.62cm]
\eprfill{1}{2}{qubitgreen} \eprfill{2}{2}{lightgreen} \eprfill{3}{2}{kerrorange}
\eprfill{1}{3}{lightgreen} \eprfill{2}{3}{lightblue}
\eprfill{1}{4}{kerrorange} \eprfill{3}{4}{lightblue}
\eprgrid
\eprlabel{0}{Sys II} \eprtext{1}{0}{\textbf{Qubit}} \eprtext{2}{0}{\textbf{Purcell Filter Res.}} \eprtext{3}{0}{\textbf{Readout Res.}}
\eprlabel{1}{Frequency} \eprtext{1}{1}{6.01 GHz} \eprtext{2}{1}{7.66 GHz} \eprtext{3}{1}{7.44 GHz}
\eprlabel{2}{Qubit} \eprtext{1}{2}{289 MHz} \eprtext{2}{2}{2.95 MHz} \eprtext{3}{2}{0.006 MHz}
\eprlabel{3}{Purcell Filter Res.} \eprtext{1}{3}{2.95 MHz} \eprtext{2}{3}{7.44 kHz} \eprtext{3}{3}{1.86e-4 MHz}
\eprlabel{4}{Readout Res.} \eprtext{1}{4}{0.006 MHz} \eprtext{2}{4}{1.86e-4 MHz} \eprtext{3}{4}{3.44 kHz}
\end{tikzpicture}
\end{table}

\begin{table}[tbp]
\caption{The table summarizes the symbols, their definitions, and their corresponding values at the design point used for all quantum-mechanical results presented in this study. Angular frequencies and decay rates are reported as ($\omega/2\pi$). The qubit--mode coupling strengths are extracted from the cross-Kerr coefficients reported in Table~\ref{tab:epr}b, following the procedure described. The parameters $g_k$ and $\kappa_r$ are treated as design variables and are kept identical across the circuits considered in Table~\ref{tab:circuits} to ensure a consistent basis for comparison. The mode frequencies and the couplings $g$ and $g_q$ come from the drawn layout of Fig.~\ref{fig:layouts}b, whose filter--readout coupling is of order 10~MHz or less (Section~\ref{sec:em}); $g_k/2\pi = 100$~MHz is therefore a design target added to the model, not a property of the drawn layout, and Fig.~\ref{fig:s21} refers to the drawn layout rather than to this design point.}
\label{tab:params}
\centering
\renewcommand{\arraystretch}{1.15}
\begin{tabular}{lll}
\hline\hline
\textbf{symbol} & \textbf{meaning} & \textbf{value}\\ \hline
$\omega_q/2\pi$ & transmon transition frequency & 6.010 GHz\\
$\alpha/2\pi$ & transmon anharmonicity & $-289$ MHz\\
$\omega_f/2\pi$ & Purcell-filter resonator frequency & 7.660 GHz\\
$\omega_r/2\pi$ & readout resonator frequency (ported) & 7.440 GHz\\
$g/2\pi$ & qubit--filter coupling & 127.8 MHz\\
$g_q/2\pi$ & direct qubit--readout coupling & 5.05 MHz\\
$g_k/2\pi$ & filter--readout coupling & 100 MHz\\
$\kappa_r/2\pi$ & external linewidth of the readout resonator & 40 MHz\\
$\kappa_f^{\mathrm{int}}/2\pi$ & internal loss rate of the filter mode & 20 kHz\\
$1/\gamma_{\mathrm{int}}$ & intrinsic (non-radiative) qubit lifetime & 60 $\mu$s\\
$1/\gamma_\phi$ & pure dephasing time & 120 $\mu$s\\
$n_{\mathrm{th}}$ & residual thermal occupation of the port & 0.02\\
$p_{\mathrm{th}}$ & residual excited-state population before a shot & 0.002\\
$\eta$ & measurement-chain efficiency & 0.70\\
$T_{\mathrm{int}}$ & integration time & 280 ns\\
$\bar{n}$ & peak photon number in either mode & 10\\
NH & Hilbert-space cutoff per resonator mode & 6\\ \hline\hline
\end{tabular}
\end{table}

\begin{table}[tbp]
\caption{The circuits compared under identical readout constraints: the same peak photon number, integration time, measurement efficiency, thermal population and intrinsic qubit lifetime, with the drive frequency separately optimized for each. A is the conventional ordering with the couplings of Table~\ref{tab:epr}a; B$_0$ is the Sys~II frequency plan with the direct path removed ($g_q = 0$), i.e.\ the bandpass-filter chain of Ref.~\cite{r7}, which by the relabelling argument of Section~\ref{sec:purcell} is also the conventional ordering redrawn with the Sys~II frequencies; B is the proposed ordering with the couplings of Table~\ref{tab:epr}b; C is the proposed ordering with the direct coupling raised to the rotating-wave interference null of Eq.~\ref{eq:null}. At the null the closed form gives zero, and the exact eigenvalue gives the internal-loss floor discussed in Section~\ref{sec:purcell} and shown in Fig.~\ref{fig:A1}. The total $T_1$, and hence $\varepsilon_{\mathrm{relax}}$, uses the exact eigenvalue, which includes the internal loss of the filter, for every column. $|\theta|$ is the mixing angle of Eq.~\ref{eq:rotation}, $\kappa_-$ and $\kappa_+$ are the port linewidths of the lower and upper normal modes, and $\bar{n}_f$ and $\bar{n}_r$ are the peak photon numbers of the qubit-coupled and ported modes.}
\label{tab:circuits}
\centering
\renewcommand{\arraystretch}{1.15}
\begin{tabular}{lcccc}
\hline\hline
\textbf{quantity}
& \begin{tabular}[t]{@{}c@{}}\textbf{A: Conventional}\\ \textbf{(Sys I)}\end{tabular}
& \begin{tabular}[t]{@{}c@{}}\textbf{B$_\mathbf{0}$: Sys II,}\\ $\boldsymbol{g_q = 0}$\\ \textbf{(bandpass chain)}\end{tabular}
& \begin{tabular}[t]{@{}c@{}}\textbf{B: Proposed}\\ \textbf{(Sys II)}\end{tabular}
& \begin{tabular}[t]{@{}c@{}}\textbf{C: at the null}\\ \textbf{(direct coupling)}\end{tabular}\\ \hline
$g_q/2\pi$ (MHz) & 4.58 & 0 & 5.05 & 7.74\\
$\Gamma_{\mathrm{P}}$, closed form (s$^{-1}$) & $2.4\times10^{2}$ & $7.4\times10^{3}$ & $9.0\times10^{2}$ & 0\\
$\Gamma_{\mathrm{P}}$, Eq.~\ref{eq:A4} incl.\ filter loss (s$^{-1}$) & $8.8\times10^{2}$ & $8.2\times10^{3}$ & $1.65\times10^{3}$ & $7.5\times10^{2}$\\
$\Gamma_{\mathrm{P}}$, exact eigenvalue (s$^{-1}$) & $8.6\times10^{2}$ & $7.9\times10^{3}$ & $1.6\times10^{3}$ & $7.4\times10^{2}$\\
Purcell-limited $T_1$, Eq.~\ref{eq:gammaP} ($\mu$s) & $4.2\times10^{3}$ & $1.3\times10^{2}$ & $1.1\times10^{3}$ & $\infty$\\
total $T_1$, exact $\Gamma_{\mathrm{P}}$ ($\mu$s) & 57.05 & 40.63 & 54.76 & 57.45\\
$|\theta|$ (deg) & 7.3 & 21.1 & 21.1 & 21.1\\
$\kappa_-/2\pi$, $\kappa_+/2\pi$ (MHz) & 0.64, 39.4 & 34.8, 5.2 & 34.8, 5.2 & 34.8, 5.2\\
$2\chi_r/2\pi$ (MHz) & $-0.076$ & $-0.776$ & $-0.698$ & $-0.660$\\
$2\chi_f/2\pi$ (MHz) & $-2.538$ & $-2.207$ & $-2.260$ & $-2.289$\\
$n_{\mathrm{crit}}$ & 50 & 50 & 49 & 48\\
peak photon number $\bar{n}_f$, $\bar{n}_r$ & 10, 0.18 & 10, 1.45 & 10, 1.44 & 10, 1.44\\
ring-up time (ns) & 12.7 & 6.0 & 6.0 & 6.0\\
SNR at $T_{\mathrm{int}} = 280$ ns & 3.59 & 6.01 & 6.11 & 6.16\\
$\varepsilon_{\mathrm{sep}}$ & $5.5\times10^{-3}$ & $1.1\times10^{-5}$ & $7.9\times10^{-6}$ & $6.5\times10^{-6}$\\
$\varepsilon_{\mathrm{relax}}$ & $5.1\times10^{-3}$ & $7.0\times10^{-3}$ & $5.2\times10^{-3}$ & $5.0\times10^{-3}$\\
$F_{|0\rangle}$ (\%) & 99.248 & 99.799 & 99.799 & 99.799\\
$F_{|1\rangle}$ (\%) & 98.936 & 99.298 & 99.478 & 99.503\\
assignment error & $9.1\times10^{-3}$ & $4.5\times10^{-3}$ & $3.6\times10^{-3}$ & $3.5\times10^{-3}$\\ \hline\hline
\end{tabular}
\end{table}

Fig.~\ref{fig:s21} shows the transmission spectra of the two drawn layouts, which do not contain the design value of $g_k$ (Section~\ref{sec:em}). In Sys~I the qubit resonance appears as a narrow dip at 6.01~GHz; in Sys~II it is suppressed below the noise floor of the simulation. The disappearance of a dip in $S_{21}$ demonstrates isolation from the port, but it does not by itself give the qubit $T_1$, and it can equally indicate that the readout signal has been lost. Fig.~\ref{fig:s21} is therefore presented only as a consistency check on the layout: it shows that the filter is transparent at the readout frequency and opaque at the qubit frequency, which is necessary but not sufficient. The quantitative Purcell statement is made through $\operatorname{Re}Y(\omega_q)$ and the resulting $T_1$, in Fig.~\ref{fig:admittance} and Table~\ref{tab:circuits}, and the quantitative readout statement through the error budget of Section~\ref{sec:readout}.

\begin{table}[tbp]
\caption{Parameters and results of the single-shot I/Q calculation behind Figs.~\ref{fig:iq} and \ref{fig:A2}, at the design point of Table~\ref{tab:params} for circuits A and B of Table~\ref{tab:circuits}. The noise scale is not an input: it is fixed by the first-principles SNR of Eq.~\ref{eq:snr}.}
\label{tab:iq}
\centering
\renewcommand{\arraystretch}{1.15}
\begin{tabular}{lcc}
\hline\hline
\textbf{parameter} & \textbf{A: Conventional (Sys I)} & \textbf{B: Proposed (Sys II)}\\ \hline
couplings & Table~\ref{tab:epr}a & Table~\ref{tab:epr}b\\
drive frequency $\omega_d/2\pi$ (GHz) & 7.877 (lower normal mode) & 7.699 (upper normal mode)\\
peak photon number $\bar{n}_f$, $\bar{n}_r$ & 10, 0.18 & 10, 1.44\\
integration time $T_{\mathrm{int}}$; efficiency $\eta$ & 280 ns; 0.70 & 280 ns; 0.70\\
SNR, Eq.~\ref{eq:snr} & 3.59 & 6.11\\
$\kappa_{\mathrm{SNR}} = \sigma/|\mu_1 - \mu_0| = 1/(\sqrt{2}\,\mathrm{SNR})$ & 0.197 & 0.116\\
single shots per state / random seed & 3200 / 12345 & 3200 / 12345\\
$F_s$, Eq.~\ref{eq:Fs} at the SNR of Eq.~\ref{eq:snr} ($=1-\varepsilon_{\mathrm{sep}}$) & 99.45\% & 99.9992\%\\
$F_s$, Gaussian fits to the shots & 99.51\% & 99.9994\%\\
misassigned shots, $|0\rangle$ / $|1\rangle$ & 19 / 19 & 0 / 0\\ \hline\hline
\end{tabular}
\end{table}

\subsection{Readout simulation and single-shot error budget}
\label{sec:readout}

This section computes the single-shot performance of the circuits of Table~\ref{tab:circuits} from first principles: the signal-to-noise ratio is obtained from an input--output calculation, Eq.~\ref{eq:snr}, and is not assumed anywhere. It begins with the separation fidelity at a given signal-to-noise ratio. The separation fidelity $F_s$ measures the distinguishability of the two Gaussian-fitted I/Q distributions and nothing else. It is not the readout fidelity, which is an end-to-end quantity that also contains relaxation during the measurement, residual thermal population, amplifier noise and measurement efficiency; these enter through Eqs.~\ref{eq:snr} and \ref{eq:budget} below. For two Gaussian clouds of equal isotropic variance $\sigma^2$ separated by $\Delta\mu$, the optimal linear classifier threshold lies at the midpoint~\cite{r5,r29} expressed as:
\begin{equation}
F_s = 1 - \tfrac{1}{2}\operatorname{erfc}\left(\frac{\Delta\mu}{2\sqrt{2}\,\sigma}\right),\qquad
\mathrm{SNR}_{\mathrm{IQ}} \equiv \frac{\Delta\mu}{\sigma} = \frac{1}{\kappa_{\mathrm{SNR}}}
\label{eq:Fs}
\end{equation}
Eq.~\ref{eq:Fs} gives the separation fidelity for a given noise scale $\sigma$; what the circuit determines is that noise scale relative to the pointer separation, and it is fixed by Eq.~\ref{eq:snr} below. Fig.~\ref{fig:iq} therefore shows single-shot I/Q distributions of circuits A and B at the design point, with $\sigma$ set by the first-principles SNR rather than assumed. The mean vectors $\mu_0$ and $\mu_1$ are the matched-filter integrals, over $T_{\mathrm{int}} = 280$~ns, of the ported-mode pointer trajectories $\alpha_g(t)$ and $\alpha_e(t)$ obtained from the qubit-state-conditioned Langevin equations, Eq.~\ref{eq:langevin} (for these linear resonator dynamics they coincide with $\langle a(t)\rangle$ from the master equation), and each shot adds Gaussian noise of standard deviation $\sigma$ such that $|\mu_1 - \mu_0|/\sigma = \sqrt{2}\,\mathrm{SNR}$, the factor $\sqrt{2}$ converting between the conventions of Eqs.~\ref{eq:Fs} and \ref{eq:snr}. Adding classical Gaussian noise to the integrated quadrature is a legitimate model of the vacuum noise and of the noise added by the amplification chain, whose efficiency $\eta$ enters through Eq.~\ref{eq:snr}, but it does not reproduce the quantum single-shot variability, which would require stochastic trajectories. For circuit A ($\mathrm{SNR} = 3.59$, $\kappa_{\mathrm{SNR}} = \sigma/|\mu_1 - \mu_0| = 0.197$) the two clouds overlap visibly, 19 of the 3200 shots per state are misassigned, and the Gaussian fits give $F_s = 99.51\%$, against $1 - \varepsilon_{\mathrm{sep}} = 99.45\%$ from Eqs.~\ref{eq:Fs} and \ref{eq:snr}; for circuit B ($\mathrm{SNR} = 6.11$, $\kappa_{\mathrm{SNR}} = 0.116$) the clouds are $8.6\sigma$ apart, no shot is misassigned, and $F_s = 99.9994\%$ (99.9992\%). Fig.~\ref{fig:iq} thus displays directly the pointer-state separation delivered by each circuit, and the gain of circuit B is visible without reference to any assumed noise level. Relaxation during the measurement and the residual thermal population are not included in the shots; they are added through Eq.~\ref{eq:budget} and give the $F_{|0\rangle}$ and $F_{|1\rangle}$ of Table~\ref{tab:circuits}. The parameters are listed in Table~\ref{tab:iq}. Fig.~\ref{fig:A2} shows the dependence of $F_{|0\rangle}$ and $F_{|1\rangle}$ on $\kappa_{\mathrm{SNR}}$, with the operating points of circuits A and B marked. It restates Eq.~\ref{eq:Fs} and is included because it makes the sensitivity visible; the relation between separation error and signal-to-noise ratio is standard and is not presented here as a new result.

\begin{figure}[tbp]
\centering
\includegraphics[width=0.9\linewidth]{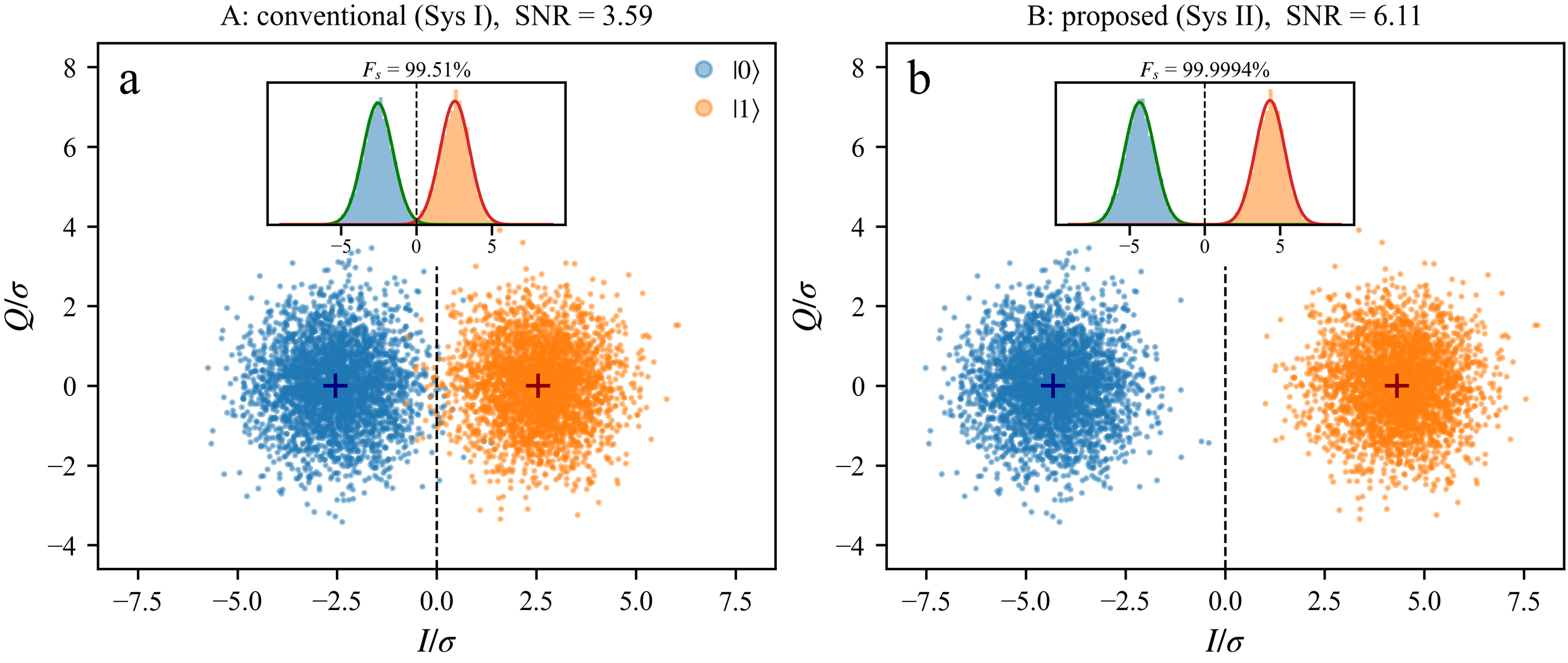}
\caption{Single-shot I/Q distributions at the design point for (a) circuit A (conventional ordering, Sys~I) and (b) circuit B (proposed ordering, Sys~II) of Table~\ref{tab:circuits}, for qubit states $|0\rangle$ (blue) and $|1\rangle$ (orange), 3200 shots each, in units of the noise standard deviation $\sigma$. The noise is not an input: $\sigma$ is fixed by the first-principles SNR of Eq.~\ref{eq:snr}, $|\mu_1 - \mu_0|/\sigma = \sqrt{2}\,\mathrm{SNR} = 5.08$ for A and 8.64 for B. Crosses: mean vectors; dashed line: optimal threshold. Insets: projections on the discriminant axis with Gaussian fits, giving $F_s = 99.51\%$ for A and 99.9994\% for B. Relaxation and thermal population are not included in the shots (see Eq.~\ref{eq:budget}); the parameters are listed in Table~\ref{tab:iq}.}
\label{fig:iq}
\end{figure}

The noise scale used in Fig.~\ref{fig:iq} follows from the input--output calculation. Driving the port and solving the qubit-state-conditioned Langevin equations, Eq.~\ref{eq:langevin}, gives the two intracavity pointer trajectories $\alpha_g(t)$ and $\alpha_e(t)$ of the ported mode, and the signal-to-noise ratio accumulated in an integration window is, following the conventions of Refs.~\cite{r5,r26} and the fast-readout analyses of Refs.~\cite{r13,r29,r50}:
\begin{equation}
\mathrm{SNR}^2 = 2\eta\kappa_r\int_0^{T_{\mathrm{int}}}\left|\alpha_e(t) - \alpha_g(t)\right|^2 dt,\qquad
\varepsilon_{\mathrm{sep}} = \tfrac{1}{2}\operatorname{erfc}\left(\mathrm{SNR}/2\right)
\label{eq:snr}
\end{equation}
where $\eta$ is the overall measurement-chain efficiency (the product of the line transmission, the quantum efficiency of the first amplifier and any subsequent losses); Eq.~\ref{eq:snr} fixes the noise scale of Figs.~\ref{fig:iq} and \ref{fig:A2}. Eq.~\ref{eq:snr} also makes explicit the only link between the circuit and the readout figure of merit: the qubit-state-dependent pull delivered to the ported mode, Eq.~\ref{eq:chi}, sets the pointer separation $|\alpha_e - \alpha_g|$, which sets the SNR and, through Eq.~\ref{eq:budget}, the assignment error. No correlation measure of Section~\ref{sec:corr} enters this chain. The complete single-shot assignment error adds the state-preparation and state-survival terms~\cite{r13,r29,r50}, introduced as:
\begin{equation}
\varepsilon_0 = \varepsilon_{\mathrm{sep}} + p_{\mathrm{th}},\qquad
\varepsilon_1 = \varepsilon_{\mathrm{sep}} + 1 - e^{-(T_{\mathrm{int}} + T_{\mathrm{ring}})/T_1},\qquad
F = 1 - \tfrac{1}{2}(\varepsilon_0 + \varepsilon_1),\qquad
T_1 = (\Gamma_{\mathrm{P}} + \gamma_{\mathrm{int}})^{-1}
\label{eq:budget}
\end{equation}
where $p_{\mathrm{th}}$ is the residual excited-state population before the shot and $T_{\mathrm{ring}}$ is the time for the pointer separation to reach $1 - 1/e$ of its steady value. Eq.~\ref{eq:budget} counts explicitly the unwanted state transitions that occur during the measurement, which the signal-to-noise ratio of Eq.~\ref{eq:snr} does not describe. Quantum non-demolition (QND) fidelity is not computed, because it requires a stochastic-trajectory treatment of the measurement back-action. The circuits of Table~\ref{tab:circuits} are compared under identical constraints: the same peak photon number in either mode ($\bar{n} = 10$, to be compared with $n_{\mathrm{crit}} = 49$), the same integration time, the same measurement efficiency $\eta = 0.70$, the same residual thermal population $p_{\mathrm{th}} = 0.002$ and the same intrinsic qubit lifetime $1/\gamma_{\mathrm{int}} = 60$~$\mu$s. The drive frequency is separately optimized for each circuit, because the hybridized resonance is not at the bare mode frequency. Circuit A is the conventional ordering with the couplings of Table~\ref{tab:epr}a; circuit B is the proposed ordering with the couplings of Table~\ref{tab:epr}b; circuit C is the proposed ordering with the direct coupling raised to the interference null of Eq.~\ref{eq:null}. Table~\ref{tab:circuits} collects the results at $T_{\mathrm{int}} = 280$~ns and Fig.~\ref{fig:A4} shows the assignment error against integration time. Circuits A and B give assignment errors of $9.1\times10^{-3}$ and $3.6\times10^{-3}$. The improvement comes from the frequency plan of Sys~II, whose two modes are 220~MHz apart and therefore hybridize strongly enough to deliver the qubit pull to the port ($2\chi_r/2\pi = -0.70$~MHz against $-0.08$~MHz for Sys~I). Circuit C reaches $3.5\times10^{-3}$, a marginal further gain, because at this operating point the error budget is already dominated by the intrinsic lifetime and the thermal population rather than by Purcell decay. The value of the interference condition is visible elsewhere: it removes the radiative channel entirely, so the readout bandwidth can be chosen without paying for it in $T_1$. At the design point the radiative rate of Eq.~\ref{eq:gammaP} is $9.0\times10^{2}$~s$^{-1}$ for circuit B and zero for circuit C, which retains only the internal-loss floor of $7.4\times10^{2}$~s$^{-1}$; and if $g_k$ and $\kappa_r$ are pushed to the values that minimize the error at a 40~ns integration time, the same comparison becomes $T_1 = 5.4$~$\mu$s for circuit B against 57.4~$\mu$s for circuit C, set only by the intrinsic lifetime and the internal loss of the filter. Decoupling the choice of readout bandwidth from the qubit lifetime, rather than a large improvement in assignment fidelity at one fixed operating point, is what the interference condition buys.

Two further comparisons show what each ingredient contributes. First, by the relabelling argument of Section~\ref{sec:purcell}, the conventional ordering redrawn with the Sys~II frequency plan is simply circuit B with the direct path removed ($g_q = 0$), i.e.\ the bandpass-filter chain of Ref.~\cite{r7}. For that circuit (column B$_0$ of Table~\ref{tab:circuits}) the exact eigenvalue gives $\Gamma_{\mathrm{P}} = 7.9\times10^{3}$~s$^{-1}$ and $T_1 = 40.6$~$\mu$s, against $1.6\times10^{3}$~s$^{-1}$ ($T_1 = 54.8$~$\mu$s) for circuit B and $7.4\times10^{2}$~s$^{-1}$ ($T_1 = 57.5$~$\mu$s) for circuit C, all including the internal loss of the filter. At a fixed frequency plan, therefore, the direct path already present in the layout suppresses the radiative rate fivefold relative to a standard bandpass filter, and tuning it to the null of Eq.~\ref{eq:null} removes the radiative channel, leaving only the internal-loss floor. Because B$_0$ and B have nearly the same SNR (6.01 and 6.11), the direct path also lowers the assignment error at this frequency plan, from $4.5\times10^{-3}$ to $3.6\times10^{-3}$, through $T_1$ alone. Second, because $\mathrm{SNR}^2$ in Eq.~\ref{eq:snr} is linear in $\eta$ while $\varepsilon_{\mathrm{relax}}$ and $p_{\mathrm{th}}$ do not depend on it, the dependence on the measurement-chain efficiency follows directly from Table~\ref{tab:circuits}. At $\eta = 0.3$ the SNR falls to 2.35 for A and 4.00 for B, and the assignment error rises to $5.2\times10^{-2}$ and $6.0\times10^{-3}$, respectively; at $\eta = 0.9$ it is $5.5\times10^{-3}$ and $3.6\times10^{-3}$. The advantage of the Sys~II frequency plan therefore grows as the efficiency of the chain decreases, because circuit A is limited by separation while circuit B remains limited by relaxation.

\begin{figure}[tbp]
\centering
\includegraphics[width=0.7\linewidth]{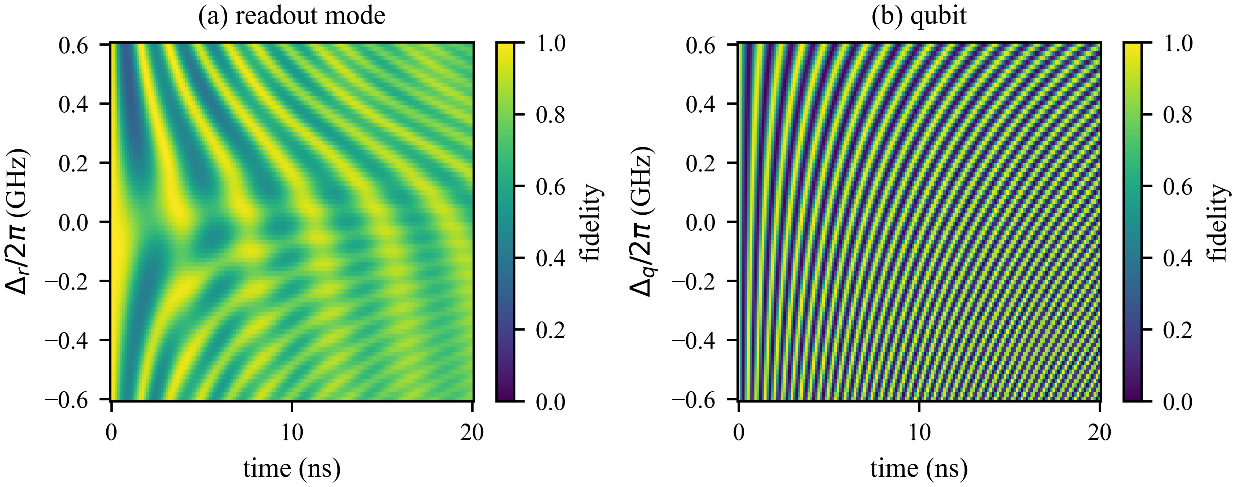}
\caption{Fidelity of (a) the readout-mode state and (b) the qubit state with respect to their initial values, as a function of time and detuning ($\Delta_r$ and $\Delta_q$ respectively), for the proposed circuit at the design point of Table~\ref{tab:params}. The fidelity is $F(\rho,\sigma) = \bigl[\operatorname{Tr}\sqrt{\sqrt{\sigma}\,\rho\,\sqrt{\sigma}}\bigr]^2$~\cite{r18,r25}, evaluated on the reduced state of each subsystem.}
\label{fig:fidelity}
\end{figure}

Fig.~\ref{fig:A3} and Fig.~\ref{fig:fidelity} show the state-level behaviour behind these numbers. Fig.~\ref{fig:A3} gives the initial and final Wigner functions of the qubit and of the readout mode for the two architectures, computed with the transmon treated as a three-level Duffing oscillator initialised in $(|0\rangle + |1\rangle)/\sqrt{2}$ and with the one-sided port dissipators of Eq.~\ref{eq:master}. Fig.~\ref{fig:fidelity} shows the fidelity of the readout-mode and of the qubit state with respect to their initial values, as a function of time and of detuning. The degradation at small detuning is the expected consequence of the qubit and the modes approaching resonance, where the dispersive treatment of Section~\ref{sec:dispersive} no longer applies. Convergence of all master-equation results with the per-mode Hilbert-space cutoff is shown in Fig.~\ref{fig:A5}; the quantities quoted in Table~\ref{tab:circuits} change by less than 1\% between NH = 6 and NH = 8.

\subsection{Correlations between the readout and filter modes}
\label{sec:corr}

Although the correlation between the filter mode and the readout mode is not conventionally considered a readout figure of merit, comparing the two systems from this perspective can nevertheless provide useful additional insight. The two internal modes are strongly hybridized, so it is worth establishing quantitatively, and with an explicitly stated convention, which correlations such a circuit produces. The measures below are therefore reported as a diagnostic of the model rather than as a readout resource. The nonclassicality at issue here is Gaussian: the resonator Hamiltonian is quadratic, so neither mode acquires Wigner negativity, and the principal term of Eq.~\ref{eq:H} that can generate squeezing is the counter-rotating part of the filter--readout coupling, $g_k(ab + a^+b^+)$, which is a two-mode-squeezing interaction. This term, absent from Eq.~\ref{eq:Hrwa}, produces the transient entanglement of Fig.~\ref{fig:transient}. The correlation measures used below are the standard continuous-variable ones~\cite{r16,r19,r20,r21,r35,r36}. The quadratures are $x = (a + a^+)/\sqrt{2}$ and $p = -i(a - a^+)/\sqrt{2}$, the covariance matrix is $\sigma_{ij} = \langle\{\Delta u_i, \Delta u_j\}\rangle/2$ with $\Delta u = u - \langle u\rangle$ in the ordering $(x_r, p_r, x_f, p_f)$. With this convention a covariance matrix describes a physical state only if:
\begin{equation}
\sigma + \frac{i}{2}\Omega \geq 0 \quad\Leftrightarrow\quad \nu_{1,2} \geq \tfrac{1}{2},\qquad
\Omega = \bigoplus_{j=1,2}\begin{pmatrix} 0 & 1\\ -1 & 0 \end{pmatrix}
\label{eq:physical}
\end{equation}
where $\nu_{1,2}$ are the Symplectic eigenvalues, the moduli of the eigenvalues of $i\Omega\sigma$. Every covariance matrix used below is tested against Eq.~\ref{eq:physical} before any correlation measure is evaluated.

For Gaussian states the Peres--Horodecki criterion in Simon's form~\cite{r20} is a statement about the partial transpose, which is the mirror reflection $p_f \to -p_f$ introduced as:
\begin{equation}
\tilde{\nu}_- < \tfrac{1}{2} \quad\Leftrightarrow\quad \text{entangled},\qquad
E_{\mathcal{N}} = \max\left\{0, -\log_2(2\tilde{\nu}_-)\right\}
\label{eq:ppt}
\end{equation}
with $\tilde{\nu}_-$ the smaller Symplectic eigenvalue of the partially transposed matrix. The logarithmic negativity quoted below follows from $E_{\mathcal{N}} = \max[0, -\log_2(2\tilde{\nu}_-)]$~\cite{r47}. Gaussian quantum discord, with the filter mode measured~\cite{r35,r36} is expressed as:
\begin{equation}
\mathcal{D} = f(\sqrt{B}) - f(\nu_-) - f(\nu_+) + f\bigl(\sqrt{E_{\mathrm{min}}}\bigr),\qquad
f(x) = \frac{x+1}{2}\log_2\frac{x+1}{2} - \frac{x-1}{2}\log_2\frac{x-1}{2}
\label{eq:discord}
\end{equation}
where $B$ is the determinant of the filter block, $\nu_\mp$ are the Symplectic eigenvalues of the full matrix, and $E_{\mathrm{min}}$ is the minimum, over all single-mode Gaussian measurements, of the determinant of the conditional covariance matrix of the readout mode. The implementation is verified against states whose correlations are known analytically: for a two-mode squeezed vacuum of squeezing parameter $r$ it returns $E_{\mathcal{N}} = 2r/\ln 2$ and $\mathcal{D} = I/2$ at $r = 0.2$, 0.5, 1.0 and 1.5, where $I$ is the mutual information; it returns zero for product states; and it satisfies $0 \le \mathcal{D} \le I$ by construction. The checks are listed in Appendix~\ref{app:D}. Finally, another correlation metric, referred to as normalized amplitude correlation, is introduced and discussed in this study. It is expressed as follows:
\begin{equation}
\mathcal{E}_e = |\langle ab\rangle|\Big/\sqrt{\langle a^+a\rangle\langle b^+b\rangle}
\label{eq:ampcorr}
\end{equation}
This quantity is not an entanglement monotone and it has no threshold at unity: it can be large for a separable state and small for an entangled one. It is shown in Figs.~\ref{fig:transient}, \ref{fig:cuts} and 12 only as a plotted comparison, so that the reader can see directly that it does not track the entanglement criterion of Eq.~\ref{eq:ppt}.

\section{Results and Discussions}
\label{sec:results}

\begin{figure}[tbp]
\centering
\includegraphics[width=0.92\linewidth]{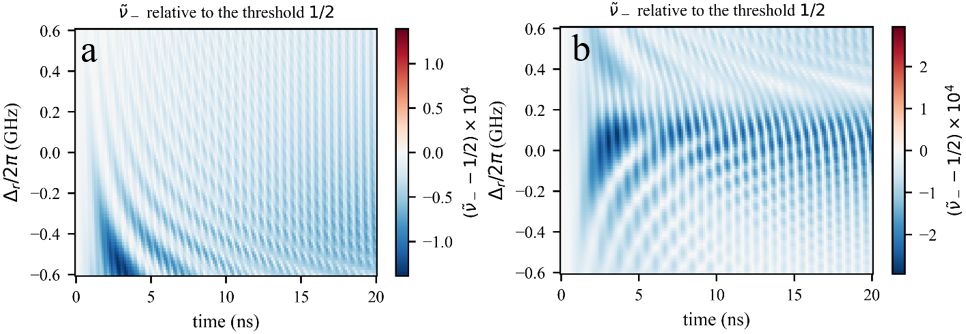}
\caption{Smallest symplectic eigenvalue $\tilde{\nu}_-$ of the partially transposed covariance matrix of the readout and filter modes (a: Sys~I and b: Sys~II), plotted as its deviation from the entanglement threshold, $(\tilde{\nu}_- - \tfrac{1}{2})\times10^{4}$, against the readout detuning $\Delta_r$ and time. By Eq.~\ref{eq:ppt} the two modes are entangled only where this quantity is negative. For Sys~I, the deviation reaches $-1.4\times10^{-4}$ and the corresponding logarithmic negativity $4.0\times10^{-4}$ bits; For Sys~II, it is negative over most of the map but never by more than $2.9\times10^{-4}$, which corresponds to a logarithmic negativity of at most $8.5\times10^{-4}$ bits and falls to $5.2\times10^{-4}$ bits as the Hilbert-space cutoff is raised from NH = 6 to NH = 8. The colour scale therefore spans the numerical resolution of the calculation, not a physically useful entanglement. Both panels use $g_k/2\pi = 100$~MHz and $\kappa_r/2\pi = 40$~MHz.}
\label{fig:nutilde}
\end{figure}

\begin{figure}[tbp]
\centering
\includegraphics[width=\linewidth]{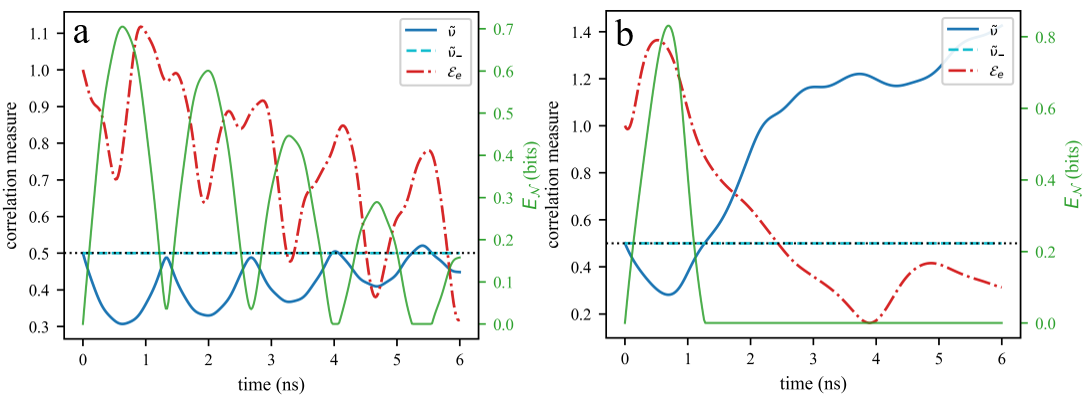}
\caption{Time evolution of the correlation measures (a: Sys~I and b: Sys~II) for the resonator and filter modes at the design point of Table~\ref{tab:params}. Solid blue: $\tilde{\nu}_-$ from the full Hamiltonian, Eq.~\ref{eq:H}; dashed cyan: $\tilde{\nu}_-$ within the rotating-wave approximation, Eq.~\ref{eq:Hrwa}; dash-dotted red: the normalised amplitude correlation $\mathcal{E}_e$ of Eq.~\ref{eq:ampcorr}, shown here for comparison only; green, right axis: logarithmic negativity $E_{\mathcal{N}}$. The dotted line is the entanglement threshold $\tilde{\nu}_- = \tfrac{1}{2}$. Entanglement is produced entirely by the counter-rotating terms in both layouts, and $\mathcal{E}_e$ tracks it in neither: in Sys~II it is a single transient that is over by 1.26~ns, whereas in Sys~I it revives at 2.00, 3.28 and 4.68~ns, because one normal mode of that layout carries only 1.6\% of $\kappa_r$.}
\label{fig:transient}
\end{figure}

In the following, the result of the simulations and the discussions relevant are introduced. Fig.~\ref{fig:nutilde} shows $\tilde{\nu}_-$ for the readout and filter modes as a function of the readout detuning and time, and in line with, Fig.~\ref{fig:transient} shows its time evolution at the design point, together with $E_{\mathcal{N}}$ and with the amplitude correlation $\mathcal{E}_e$. These figures are used to compare the two systems. In the model of Eq.~\ref{eq:H} the mode the qubit is coupled to is $b$ and the ported mode is $a$, so that for Sys~II these are the physical Purcell filter at 7.66~GHz and the physical readout resonator at 7.44~GHz, while for Sys~I they are the physical readout resonator at 7.89~GHz and the physical Purcell filter at 8.66~GHz. That exchange of names is precisely the relabelling identified below Eq.~\ref{eq:gammaP}, and it is what allows both layouts to be treated within one model and compared panel by panel. Every readout constraint is held fixed between the panels---$g_k/2\pi = 100$~MHz, $\kappa_r/2\pi = 40$~MHz, $\bar{n} = 10$ and NH = 6---so that the only difference between them is the frequency plan and the qubit--mode cross-Kerr values that the electromagnetic analysis returns for the two layouts in Table~\ref{tab:epr}. Within the rotating-wave approximation the two internal modes are coupled by a beam-splitter interaction, $g_k(a^+b + ab^+)$. A beam splitter cannot entangle inputs that are individually classical, so no entanglement is expected from a coherent or a thermal input, and none is found in any useful amount. Within this approximation the two systems behave alike, as the beam-splitter argument requires. In both the deviation $\tilde{\nu}_- - \tfrac{1}{2}$ is negative over 97\% of the map: it reaches $-1.4\times10^{-4}$ in Sys~I and $-2.9\times10^{-4}$ in Sys~II, corresponding to logarithmic negativities of at most $4.0\times10^{-4}$ and $8.5\times10^{-4}$ bits. Both residues lie at the level set by the Hilbert-space truncation, so the factor of two between them is not a physical difference in entanglement. What does differ is where the residue sits in the detuning scan. For Sys~II it is concentrated within about $\pm200$~MHz of the drive resonance and falls by a factor of three to five at the edges of the $\pm600$~MHz window, because the normal-mode splitting of that layout, $\sqrt{(\omega_f - \omega_r)^2 + 4g_k^2}/2\pi = 297$~MHz, lies well inside the scanned range. For Sys~I the two internal modes are 770~MHz apart, the splitting is 796~MHz, and the residue instead deepens monotonically towards negative detuning, from $-0.33\times10^{-4}$ at $\Delta_r/2\pi = +0.6$~GHz to $-1.34\times10^{-4}$ at $-0.6$~GHz, as the scan approaches without reaching the degeneracy at $\Delta_r/2\pi = -0.77$~GHz. Fig.~\ref{fig:nutilde} therefore shows the same physics in both architectures, sampled over different fractions of the same hybridisation curve, and neither ordering produces steady-state entanglement in any useful amount. Fig.~\ref{fig:transient} is where the two architectures behave qualitatively differently, and it is the most informative of the new comparisons. In both, the rotating-wave curve stays flat at $\tilde{\nu}_- = \tfrac{1}{2}$ to within $3\times10^{-4}$, so in both the correlation is produced entirely by the counter-rotating terms of Eq.~\ref{eq:H} and the mechanism is the same. In Sys~II it is a single event: $\tilde{\nu}_-$ falls to 0.281 at 0.70~ns, $E_{\mathcal{N}}$ peaks at 0.83 bits, and by 1.26~ns the logarithmic negativity has returned to zero while $\tilde{\nu}_-$ has risen through the threshold to a plateau at 1.27, that is, to a state separable by a wide margin. In Sys~I the first dip is followed by a train of revivals: $\tilde{\nu}_-$ returns below $\tfrac{1}{2}$ at 2.00, 3.28 and 4.68~ns with $E_{\mathcal{N}} = 0.60$, 0.45 and 0.29 bits, and it stays below the threshold for 93\% of the 6~ns record against 21\% for Sys~II. The difference follows from the rotation of Eq.~\ref{eq:rotation}, which distributes the port linewidth over the two normal modes as $\sin^2\theta$ and $\cos^2\theta$: with the mixing angles of Eq.~\ref{eq:rotation} listed in Table~\ref{tab:circuits} ($|\theta| = 7.3^\circ$ for Sys~I and $21.1^\circ$ for Sys~II) the normal modes of Sys~I are damped at 39.4 and 0.64~MHz and those of Sys~II at 34.8 and 5.2~MHz. Sys~I therefore keeps one normal mode that carries only 1.6\% of $\kappa_r$ and decays in about 250~ns; whatever counter-rotating excitation lands in it is not removed within the plotted window, and it beats against the ported mode with a period of about 1.35~ns, close to the 1.26~ns set by the Sys~I normal-mode splitting. In Sys~II the least damped normal mode is eight times broader and the transient is over within about a nanosecond. The same asymmetry appears in the ring-up times of Table~\ref{tab:circuits}, 12.7~ns for Sys~I against 6.0~ns for Sys~II. Two qualifications apply: First, the peak values are lower bounds limited by the Hilbert-space cutoff in Sys~I as in Sys~II, so the ordering of the two peaks, 0.70 against 0.83 bits, should not be read as a converged difference; what is robust is the presence or absence of the revivals, which is a property of the damping and not of the truncation. Second, neither transient is a readout resource: both are confined to the first few nanoseconds, within the ring-up time and far short of the 280~ns integration window of Table~\ref{tab:circuits}. The comparison is reported because the weak coupling of one Sys~I normal mode to the port---the same property that starves the ported mode of qubit pull---is also what leaves internal excitation that the port does not evacuate.

\begin{figure}[tbp]
\centering
\includegraphics[width=0.96\linewidth]{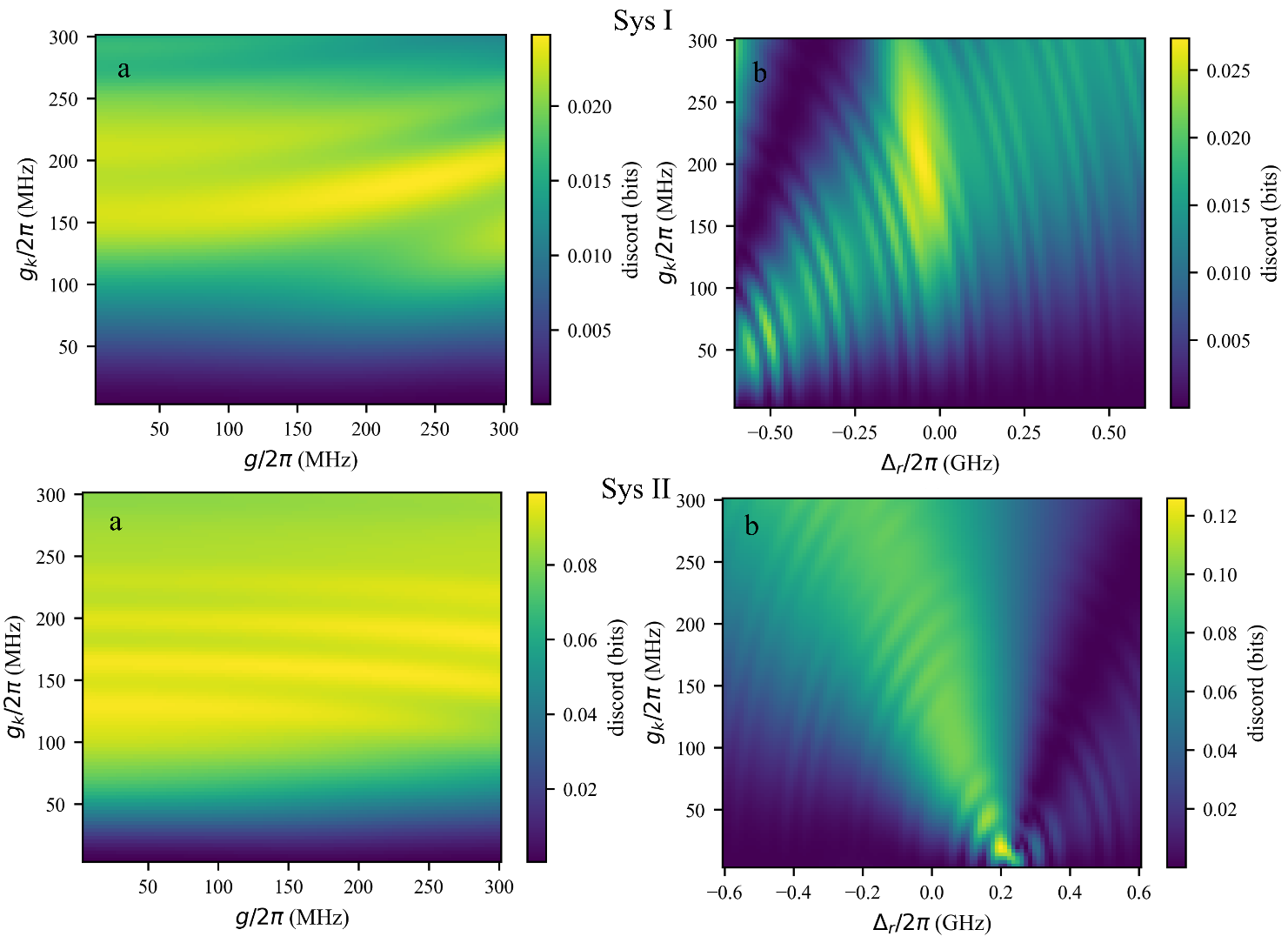}
\caption{Gaussian quantum discord of the resonator and filter modes for Sys~I (upper row) and Sys~II (lower row) in the steady state. (a) As a function of the qubit--filter coupling $g$ and the filter--readout coupling $g_k$; (b) as a function of the readout detuning $\Delta_r$ and $g_k$. The discord is evaluated from Eq.~\ref{eq:discord}; Each panel carries its own colour scale. The Sys~I discord does not exceed $2.7\times10^{-2}$ bits anywhere, against $1.3\times10^{-1}$ bits for Sys~II.}
\label{fig:discord}
\end{figure}

The Gaussian quantum discord between the resonator and filter modes in the steady state for Sys~I and Sys~II is shown in Fig.~\ref{fig:discord}. The upper row presents the corresponding maps for Sys~I, while the lower row shows the results for Sys~II. In both systems, the discord remains non-negative over the investigated parameter ranges and exhibits a similar qualitative dependence on the filter--readout coupling $g_k$. In particular, the discord increases with $g_k$ and reaches its largest values in a broad region around $g_k/2\pi \sim 50$--200~MHz. The relatively weak variation along the qubit--filter coupling $g$ in the vicinity of this ridge indicates that the resonator--filter correlation is governed predominantly by the direct resonator--filter interaction rather than by the qubit-mediated coupling. As $g_k$ is reduced toward zero, the discord also approaches zero, consistent with the fact that $g_k$ provides the direct interaction responsible for correlating the two modes considered here. The principal difference between the two layouts is therefore not the qualitative dependence on $g_k$, but the magnitude and detuning selectivity of the generated correlations. For Sys~I, the maximum discord remains below approximately $2.7\times10^{-2}$ bits over the parameter space shown, whereas Sys~II reaches approximately $1.3\times10^{-1}$ bits. Thus, the correlation between the resonator and filter modes is substantially stronger in Sys~II within the investigated parameter range. At the design point of Table~\ref{tab:params}, $g/2\pi = 127.8$~MHz and $g_k/2\pi = 100$~MHz, the discord values are approximately $1.5\times10^{-2}$ and $8.9\times10^{-2}$ bits for Sys~I and Sys~II, respectively, corresponding to a difference of roughly a factor of six.

The detuning-dependent maps in Fig.~\ref{fig:discord}b provide an additional distinction between the two layouts. For Sys~II, the discord develops a pronounced maximum around $\Delta_r/2\pi = +0.08$~GHz, with the strongest correlations concentrated in a relatively well-defined region of the detuning--$g_k$ parameter space. In contrast, Sys~I exhibits a considerably broader and weaker response over the investigated detuning range, without a comparably pronounced localized maximum. This difference is consistent with the contrasting normal-mode structures discussed in Fig.~\ref{fig:nutilde}: the narrower mode separation in Sys~II produces a more readily resolved detuning-dependent correlation feature, whereas the broader separation in Sys~I results in a less selective response over the scanned range. Consequently, Fig.~\ref{fig:discord} indicates that both layouts generate finite Gaussian quantum discord between the resonator and filter modes, but the magnitude and spectral selectivity of this correlation depend strongly on the underlying layout. Sys~II produces substantially larger discord and a more pronounced detuning-dependent feature, whereas Sys~I exhibits weaker and broader correlations. Nevertheless, the discord values in both systems remain well below one bit throughout the investigated parameter space, and neither map exhibits a sharply localized enhancement associated with a particular coupling ratio.

\begin{figure}[tbp]
\centering
\includegraphics[width=\linewidth]{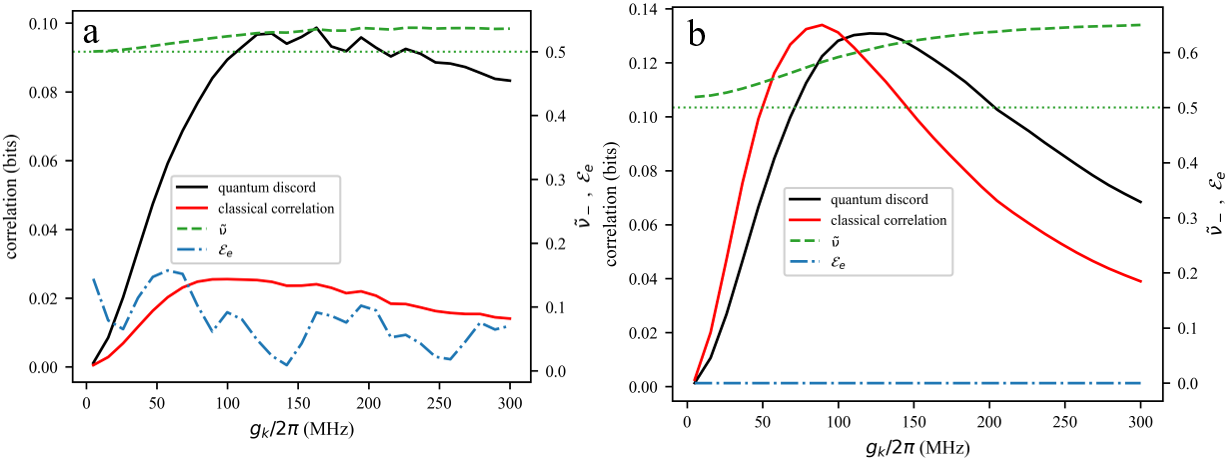}
\caption{Quantum discord (black) and classical correlation (red), left axis, and $\tilde{\nu}_-$ together with the amplitude correlation $\mathcal{E}_e$, right axis, as functions of the filter--readout coupling $g_k$, a) for a coherent initial state with $\alpha_0 = 0.5 + 0.07j$; b) thermal initial state with $n_{\mathrm{th}} = 2$. The dotted line marks $\tilde{\nu}_- = \tfrac{1}{2}$; the modes remain separable over the whole range, while $\mathcal{E}_e$ varies independently of the entanglement.}
\label{fig:cuts}
\end{figure}

Fig.~\ref{fig:cuts}a and Fig.~\ref{fig:cuts}b cuts through these maps at the design value of $g$, for a coherent and for a thermal input respectively, and add $\tilde{\nu}_-$ and $\mathcal{E}_e$ on the right-hand axis. At the design point the two internal modes have $\tilde{\nu}_- = 0.524$, so they are separable, with a quantum discord of $8.9\times10^{-2}$ bits and a classical correlation of the same order; both quantities are converged in the Hilbert-space cutoff, Fig.~\ref{fig:A5}. Over the whole range of $g_k$ the discord runs from $1.1\times10^{-3}$ to $9.9\times10^{-2}$ bits for the coherent input and from $1.3\times10^{-3}$ to $1.3\times10^{-1}$ bits for the thermal one, and $\tilde{\nu}_-$ never falls below 0.500 in either case: a thermal input carries slightly more quantum correlation than a coherent one at the same coupling, but neither produces entanglement, which is what the beam-splitter argument given above requires. The conclusion is therefore a negative one: in the operating regime the circuit produces separable internal modes carrying weak quantum correlation.

It is important to state clearly what this analysis establishes---and what it does not. First, it does not demonstrate that placing the filter before the readout resonator is intrinsically advantageous; as shown in Section~\ref{sec:purcell}, the two orderings of a purely cascaded chain are equivalent. Nor does it establish an advantage associated with the filter photon number, since the radiative decay rate in Eq.~\ref{eq:gammaP} contains no explicit dependence on photon occupation. Likewise, the analysis does not demonstrate an enhancement of nonclassicality: as shown in Section~\ref{sec:corr}, the internal modes remain separable within the operating regime considered here. Finally, no readout fidelity, QND fidelity, or experimentally measured $T_1$ is claimed, since these quantities require either experimental characterization or, in the case of QND fidelity, an appropriate stochastic-trajectory analysis. The principal result is instead the derivation of a closed-form expression for the radiative decay rate of a three-mode chain with a direct qubit-to-port pathway, given by Eq.~\ref{eq:gammaP}. The expression reproduces the two established limiting cases, remains finite in the vacuum limit, vanishes when the qubit is decoupled, and is independently validated by four numerical routes (Fig.~\ref{fig:A1}) as well as by the corresponding classical circuit admittance. In addition, Eq.~\ref{eq:null} provides an exact cancellation condition whose practical limitation is set primarily by the internal quality factor of the filter. The EPR analysis summarized in Table~\ref{tab:epr} further indicates that the layout in Fig.~\ref{fig:layout} already provides approximately 65\% of the direct coupling required by the rotating-wave cancellation condition, and about 95\% of that required by the exact solution of the lumped circuit (Fig.~\ref{fig:admittance}), suggesting that the target notch could be approached through a modest adjustment of a single coupling capacitance. The underlying physical mechanism is not itself new. Interferometric Purcell suppression has previously been demonstrated experimentally by Yen et al.~\cite{r34} and Spring et al.~\cite{r33}. The contribution of the present analysis is therefore the formulation of the general three-mode cancellation condition, its equivalent admittance representation including the full-circuit correction factor $F$, and the identification of the relevant design target in terms of a quantity that can be obtained directly from standard EPR analysis. The comparison with previous work should be interpreted in this context. The combined readout/filter architecture of Bronn et al.~\cite{r15} and the intrinsic-filter implementations of Sunada et al.~\cite{r30,r32} provide Purcell protection primarily through engineering the impedance presented to the qubit, whereas the interferometric approaches of Refs.~\cite{r33,r34} suppress decay through cancellation between emission pathways. However, Eq.~\ref{eq:gammaP} unifies these two effects: the denominator captures the impedance-shaping contribution, while the numerator describes the interference between emission amplitudes. The proposed design also inherits the practical limitations associated with strongly hybridized readout structures. The two internal modes are separated by only approximately 220~MHz, making the cancellation condition sensitive to fabrication-induced variations in the coupling capacitances and, consequently, to shifts of the Purcell-suppression notch. In addition, the filter--readout coupling of the design point has still to be realized in a layout (Section~\ref{sec:em}), and doing so changes the capacitance network through which the interference condition is set. As illustrated in Fig.~\ref{fig:A1}, a 10\% deviation in $g_q$ increases the residual decay rate by approximately one order of magnitude relative to its minimum value. Thus, the achievable internal quality factor of the filter and the fabrication tolerance of the relevant coupling parameters are likely to be the principal practical factors determining whether the predicted cancellation can be realized experimentally. Therefore, the results presented here characterize the intracavity dynamics and radiative decay of the internal circuit modes. The end-to-end readout performance of a practical system also depends on the amplification chain that follows the device, including the parametric amplifier~\cite{r22,r23,r24}. The resulting system-level readout fidelity therefore lies beyond the scope of the present analysis. The same applies to readout schemes that use squeezed light, whether generated by a phase-sensitive Josephson parametric amplifier~\cite{r48}, injected into the readout cavity~\cite{r49}, or produced and recombined by Josephson parametric converters~\cite{r31}; within the present model the pointer states are coherent, and the gain from such resources would add to, rather than replace, the frequency-plan and interference effects analyzed here.

\section{Conclusions}
\label{sec:conclusions}

We have analyzed a three-mode superconducting-qubit readout circuit comprising a transmon qubit, a Purcell-filter resonator, and a port-coupled readout resonator. For a purely serial chain, exchanging the two resonators is mathematically equivalent to relabeling the two bosonic modes and therefore provides no intrinsic advantage; likewise, the photon population of the filter does not constitute an additional degree of freedom. From the qubit self-energy, Eq.~\ref{eq:gammaP}, we derive the radiative decay rate, where $g_q$ denotes the direct qubit-to-readout coupling introduced by the physical layout in addition to the serial coupling path. The resulting expression reduces to the bandpass-filter result of Ref.~\cite{r7} when the direct path is removed and to the bare Purcell decay rate when the inter-mode coupling is set to zero. It remains finite in the vacuum limit and vanishes when the qubit is completely decoupled. Most importantly, the decay rate vanishes identically on the condition $gg_k + g_q\Delta_{qf} = 0$, which represents an exact two-path interference condition. This cancellation was verified against four independent numerical calculations and through the real part of the admittance presented to the Josephson junction. In practice, the residual limitation is set by internal loss in the filter mode. We applied this analysis to two layouts under matched readout constraints: Sys~I, corresponding to the conventional ordering, and Sys~II, corresponding to the proposed configuration. Within the common model, the two systems differ through the frequency plans obtained from the electromagnetic models and through the qubit--mode coupling strengths extracted from them, with internal-mode separations of 770~MHz for Sys~I and 220~MHz for Sys~II. Consequently, the observed performance differences arise from the resulting hybridization rather than from the nominal ordering of the resonators. In Sys~II, the stronger hybridization transfers a substantially larger fraction of the dispersive shift to the mode directly accessed by the readout port, giving $2\chi_r/2\pi = -0.698$~MHz compared with $-0.076$~MHz for Sys~I. This enhanced port-visible dispersive response is accompanied by a reduction in the single-shot assignment error from $9.1\times10^{-3}$ to $3.6\times10^{-3}$ and a reduction in the ring-up time from 12.7~ns to 6.0~ns. The same hybridization determines how the port-induced linewidth is distributed between the two normal modes. The corresponding linewidths are 34.8 and 5.2~MHz for Sys~II, compared with 39.4 and 0.64~MHz for Sys~I. In Sys~I, one normal mode remains nearly decoupled from the readout port, allowing its counter-rotating entanglement transient to revive three times within the 6~ns observation window. In Sys~II, by contrast, the transient decays within approximately 1.3~ns. At steady state, both layouts leave the two internal modes essentially separable and only weakly discordant, with maximum discord values of $2.7\times10^{-2}$ bits for Sys~I and $1.3\times10^{-1}$ bits for Sys~II. Thus, the correlation analysis quantifies the degree of mode hybridization but does not identify an additional correlation resource exploited by the readout process. Taken together, these results attribute the performance difference to the electromagnetic frequency plan and the associated hybridization, rather than to the ordering of the resonators itself, in agreement with the symmetry points expressed by Eq.~\ref{eq:gammaP}.

\appendix

\section{Derivation of Eq.~6}
\label{app:A}

The two resonator modes enter the qubit self-energy through the inverse of the non-Hermitian matrix of Eq.~\ref{eq:selfenergy}. With $\Delta_{qf} = \omega_q - \omega_f$ and $\Delta_{qr} = \omega_q - \omega_r$, and neglecting the internal loss of the filter for the moment,
\begin{equation}
(\omega_q\mathbf{1} - H_c)^{-1} = \frac{1}{D}\begin{pmatrix} \Delta_{qr} + i\kappa_r/2 & g_k\\ g_k & \Delta_{qf} \end{pmatrix},\qquad
D = \Delta_{qf}(\Delta_{qr} + i\kappa_r/2) - g_k^2
\label{eq:A1}
\end{equation}
so that, with $\mathbf{g} = (g, g_q)^{\mathrm{T}}$,
\begin{equation}
\Sigma(\omega_q) = \frac{g^2(\Delta_{qr} + i\kappa_r/2) + 2gg_qg_k + g_q^2\Delta_{qf}}{D} \equiv \frac{N}{D}
\label{eq:A2}
\end{equation}
The numerator $N$ has imaginary part $g^2\kappa_r/2$ and real part $g^2\Delta_{qr} + 2g\,g_q\,g_k + g_q^2\Delta_{qf}$, while $D$ has imaginary part $\Delta_{qf}\kappa_r/2$ and real part $\Delta_{qf}\Delta_{qr} - g_k^2$. Using $\operatorname{Im}(N/D) = [\operatorname{Im}(N)\operatorname{Re}(D) - \operatorname{Re}(N)\operatorname{Im}(D)]/|D|^2$ and collecting terms,
\begin{equation}
\Gamma_{\mathrm{P}} = -2\,\operatorname{Im}\frac{N}{D} = -2\,\frac{\operatorname{Im}(N)\operatorname{Re}(D) - \operatorname{Re}(N)\operatorname{Im}(D)}{|D|^2} = \frac{\kappa_r(gg_k + g_q\Delta_{qf})^2}{|D|^2}
\label{eq:A3}
\end{equation}
which is Eq.~\ref{eq:gammaP}. Every cross term cancels except the perfect square in the numerator, and it is that cancellation which makes the null exact rather than approximate. Retaining the internal loss of the filter replaces $\Delta_{qf}$ by $\Delta_{qf} + i\kappa_f^{\mathrm{int}}/2$ in $D$ and adds the corresponding channel to the numerator,
\begin{equation}
\Gamma_{\mathrm{P}} = \frac{\kappa_r(gg_k + g_q\Delta_{qf})^2 + \kappa_f^{\mathrm{int}}(g\Delta_{qr} + g_qg_k)^2}{\left|(\Delta_{qf} + i\kappa_f^{\mathrm{int}}/2)(\Delta_{qr} + i\kappa_r/2) - g_k^2\right|^2}
\label{eq:A4}
\end{equation}
Eq.~\ref{eq:A4} shows why the exact calculation saturates at the value quoted in Section~\ref{sec:purcell}: at the port null the first term vanishes and the second reduces to $(g/\Delta_{qf})^2\kappa_f^{\mathrm{int}}$ when $g_k$ and $g_q$ are small compared with the detunings, which is the number found numerically.

\section{Admittance of the lumped network}
\label{app:B}
\setcounter{equation}{4}
\renewcommand{\theequation}{A\arabic{equation}}
\renewcommand{\theHequation}{A\arabic{equation}}

The Purcell rate of a junction shunted by an admittance $Y(\omega)$ is $\Gamma_{\mathrm{P}} = \operatorname{Re}Y(\omega_q)/C_q$~\cite{r4,r38}. For the network of Fig.~\ref{fig:layout}b the junction node is coupled to the filter node through $C_{c1}$ and to the readout node through $C_{c2}$, the two resonator nodes are coupled through $C_{fr}$, and the readout node is loaded by the line. Eliminating the two resonator nodes by a Schur complement,
\begin{equation}
Y(\omega) = Y_{qq} - \mathbf{v}^{\mathrm{T}}\mathbf{M}^{-1}\mathbf{v},\qquad
\mathbf{v} = (Y_{qf}, Y_{qr})^{\mathrm{T}},\qquad
\mathbf{M} = \begin{pmatrix} Y_{ff} & Y_{fr}\\ Y_{fr} & Y_{rr} \end{pmatrix}
\label{eq:A5}
\end{equation}
and taking the real part, which is carried entirely by the line conductance at the readout node, gives
\begin{equation}
\operatorname{Re}Y(\omega) = \operatorname{Re}y_{\mathrm{line}}(\omega)\,\frac{\omega^2(\omega C_{c1}C_{fr} + C_{c2}y_f)^2}{(\omega^2C_{fr}^2 - y_f y_r)^2},\qquad
y_m = \omega C_{m\Sigma}\left(1 - \frac{\omega_m^2}{\omega^2}\right)
\label{eq:A6}
\end{equation}
whose numerator vanishes under the condition of Eq.~\ref{eq:admnull}. Comparing with Eq.~(\ref{eq:null}) shows that the quantised and the classical conditions differ by the factor $F = 2\omega_q^2/[\omega_f(\omega_q + \omega_f)] = 2\omega_f/(\omega_q + \omega_f) - 2(\omega_f - \omega_q)/\omega_f$, which combines the counter-rotating correction $2\omega_f/(\omega_q + \omega_f)$ with the electrostatic qubit--readout coupling $2gg_k/\omega_f$ that the series connection of $C_{c1}$ and $C_{fr}$ adds to that of $C_{c2}$; at the design point $F = 0.690$. Figure~\ref{fig:admittance} plots $\operatorname{Re}Y(\omega)/C_q$ for element values obtained from $g$, $g_q$ and $g_k$ through $g_{ij} = C_{ij}(\omega_i\omega_j)^{1/2}/[2(C_{i\Sigma}C_{j\Sigma})^{1/2}]$, with the port represented by a frequency-independent conductance that gives $\kappa_r$, together with the corresponding Purcell-limited lifetime. The position of the notch depends neither on the node capacitances nor on the port model.

\section{Numerical methods}
\label{app:C}

The master equation, Eq.~\ref{eq:master}, is solved with QuTiP~\cite{r17}. The transmon is truncated to three levels, which is sufficient because the readout is operated at $\bar{n} = 10 \ll n_{\mathrm{crit}} = 49$ and the second excited state is detuned by $|\alpha| = 2\pi\times289$~MHz; each resonator mode is truncated at NH = 6. The solver tolerances are atol $= 10^{-11}$ and rtol $= 10^{-9}$, with a maximum of $10^{5}$ internal steps. The frame rotates at the readout drive frequency, so only detunings enter the Hamiltonian. The collapse operators are $\sqrt{\kappa_r(1 + n_{\mathrm{th}})}\,a$, $\sqrt{\kappa_r n_{\mathrm{th}}}\,a^+$, $\sqrt{\kappa_f^{\mathrm{int}}}\,b$, $\sqrt{\gamma_{\mathrm{int}}}\,q$ and $\sqrt{2\gamma_\phi}\,q^+q$, exactly as written in Eq.~\ref{eq:master}; no additional environment mode is used, so the bath is not double counted. Second moments are averaged over the last quarter of each record before the Gaussian measures are evaluated, because the two-mode state oscillates at the mode--mode beat frequency and a single final-time sample would depend on the phase at which the record happens to stop. Convergence with NH is shown in Fig.~\ref{fig:A5}: between NH = 5 and NH = 8 the symplectic eigenvalues and the discord change by less than 1\%. That applies to the steady-state quantities computed within the rotating-wave approximation; the non-RWA transient of Fig.~\ref{fig:transient} is the one quantity in this paper that is not converged at $N_H = 6$, because the counter-rotating terms pump the modes, and its peak is quoted in Section~\ref{sec:corr} as a lower bound. The readout calculation of Section~\ref{sec:readout} does not use the master equation. The qubit-state-conditioned Langevin equations are linear and time-invariant, so the pointer trajectories follow in closed form as $z(t) = (e^{At} - 1)A^{-1}d$. This agrees with direct numerical integration of the same equations to a relative deviation below $10^{-9}$ and is what makes the design scans of Fig.~\ref{fig:A4} tractable. Because the response is linear in the drive, the amplitude that produces a given photon number is obtained by scaling rather than by root finding. The I/Q shots of Figs.~\ref{fig:iq} and \ref{fig:A2} use the same pointer trajectories, integrated with the matched filter $(\alpha_e - \alpha_g)^*$, with Gaussian noise fixed by Eq.~\ref{eq:snr}.

\section{Verification of the Purcell rate and of the Gaussian toolbox}
\label{app:D}

Five independent calculations of $\Gamma_{\mathrm{P}}$ are compared in Fig.~\ref{fig:A1}: the closed form of Eq.~\ref{eq:gammaP}; the port-amplitude form, in which the qubit-to-port transition amplitude is computed by solving the two-mode linear system and $\Gamma_{\mathrm{P}}$ is $\kappa_r$ times its squared modulus; the imaginary part of the complex eigenvalue of the $3\times3$ non-Hermitian single-excitation matrix, which makes no expansion in the couplings; the closed form evaluated at the Lamb-shifted qubit frequency; and a Lindblad simulation in which $|e,0,0\rangle$ is prepared with the drive off and the excited-state population is fitted to an exponential. Away from the notch, Eq.~\ref{eq:A4} agrees with the exact eigenvalue to within 4\% for all circuits of Table~\ref{tab:circuits}, and Eq.~\ref{eq:gammaP} agrees to within 6\% with the exact eigenvalue evaluated at $\kappa_f^{\mathrm{int}} = 0$, the residual difference being of higher order in $g/\Delta_{qf} = 0.077$. The Gaussian toolbox is verified against states whose correlations are known in closed form. For a two-mode squeezed vacuum of squeezing parameter $r$ it returns $E_{\mathcal{N}} = 2r/\ln 2$ at $r = 0.2$, 0.5, 1.0 and 1.5, and a discord equal to half the mutual information, which is the exact result for a pure state; for product states, thermal or vacuum, it returns exactly zero; for a classically correlated separable state it returns $E_{\mathcal{N}} = 0$ with a positive discord; and it rejects covariance matrices that violate Eq.~\ref{eq:physical}.

\setcounter{figure}{0}
\renewcommand{\thefigure}{A\arabic{figure}}
\renewcommand{\theHfigure}{A\arabic{figure}}

\begin{figure}[tbp]
\centering
\includegraphics[width=0.72\linewidth]{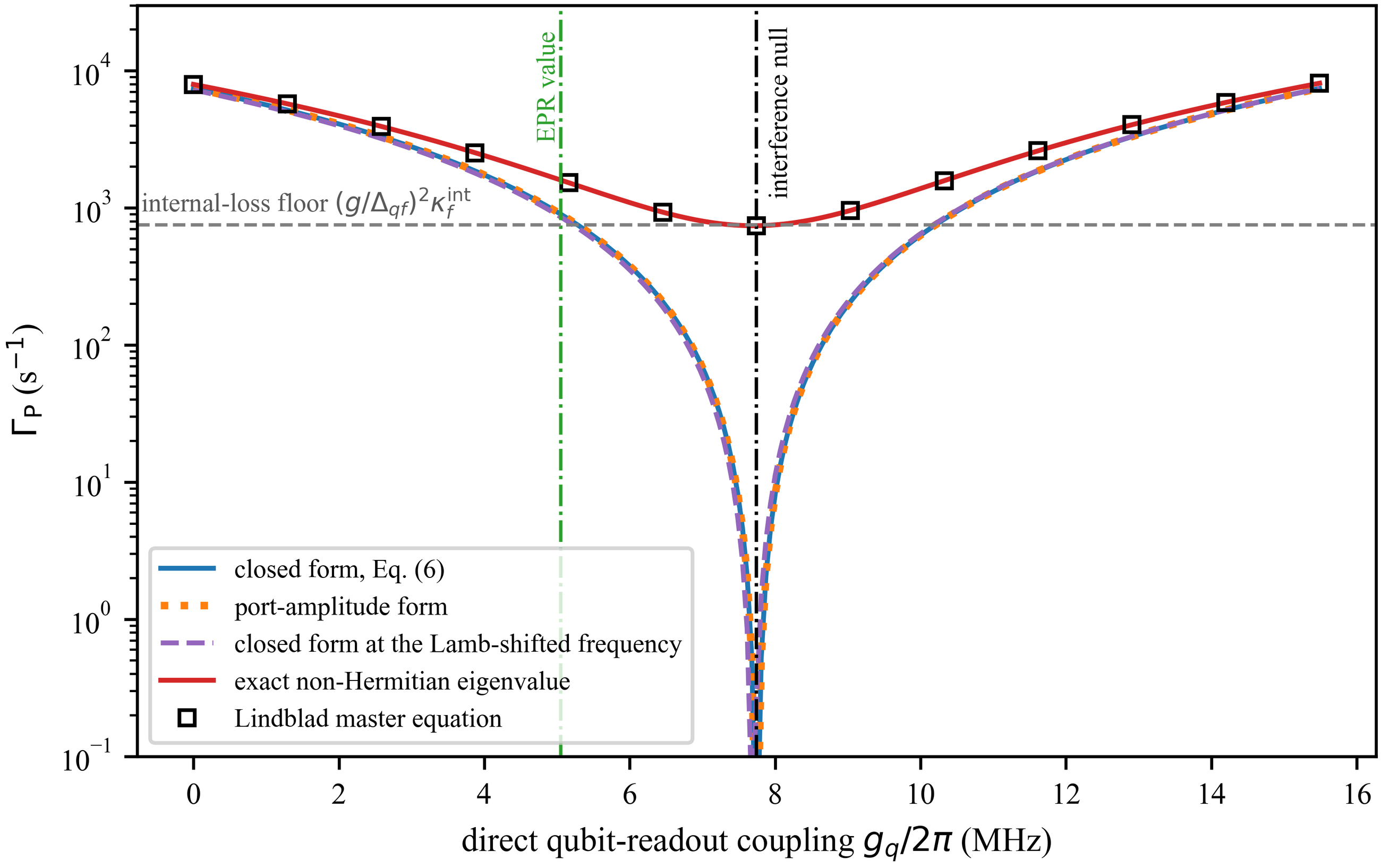}
\caption{Radiative decay rate as a function of the direct qubit--readout coupling $g_q$, at the design point of Table~\ref{tab:params}, computed by five independent routes. The vertical dash-dotted lines mark the value extracted from the energy-participation analysis of Table~\ref{tab:epr}b and the null of Eq.~\ref{eq:null}. The horizontal dashed line is the internal-loss floor $(g/\Delta_{qf})^2\kappa_f^{\mathrm{int}}$, which is what the exact calculation saturates at; with the filter loss set to zero the exact rate at the null falls to 0.26~s$^{-1}$.}
\label{fig:A1}
\end{figure}

\begin{figure}[tbp]
\centering
\includegraphics[width=0.63\linewidth]{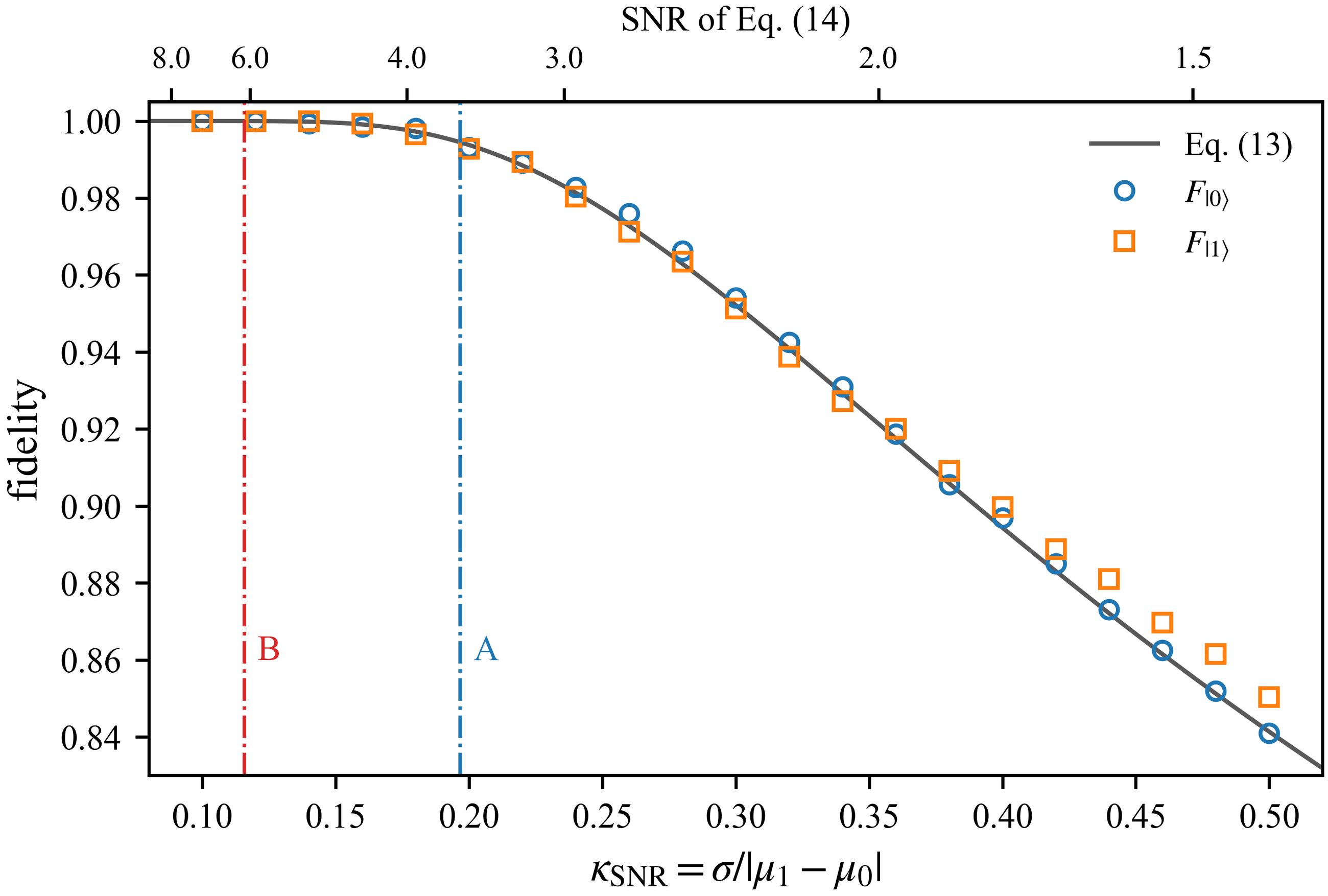}
\caption{Dependence of the separation fidelities $F_{|0\rangle}$ (blue circles) and $F_{|1\rangle}$ (orange squares) on the normalised noise parameter $\kappa_{\mathrm{SNR}}$, which sets the width of the I/Q distributions relative to their separation. Symbols: 3200 simulated shots per state; line: Eq.~\ref{eq:Fs}; vertical lines: the operating points of circuits A ($\kappa_{\mathrm{SNR}} = 0.197$) and B ($\kappa_{\mathrm{SNR}} = 0.116$) fixed by Eq.~\ref{eq:snr}. The top axis gives the corresponding SNR of Eq.~\ref{eq:snr}, $\mathrm{SNR} = 1/(\sqrt{2}\,\kappa_{\mathrm{SNR}})$.}
\label{fig:A2}
\end{figure}

\begin{figure}[tbp]
\centering
\includegraphics[width=0.65\linewidth]{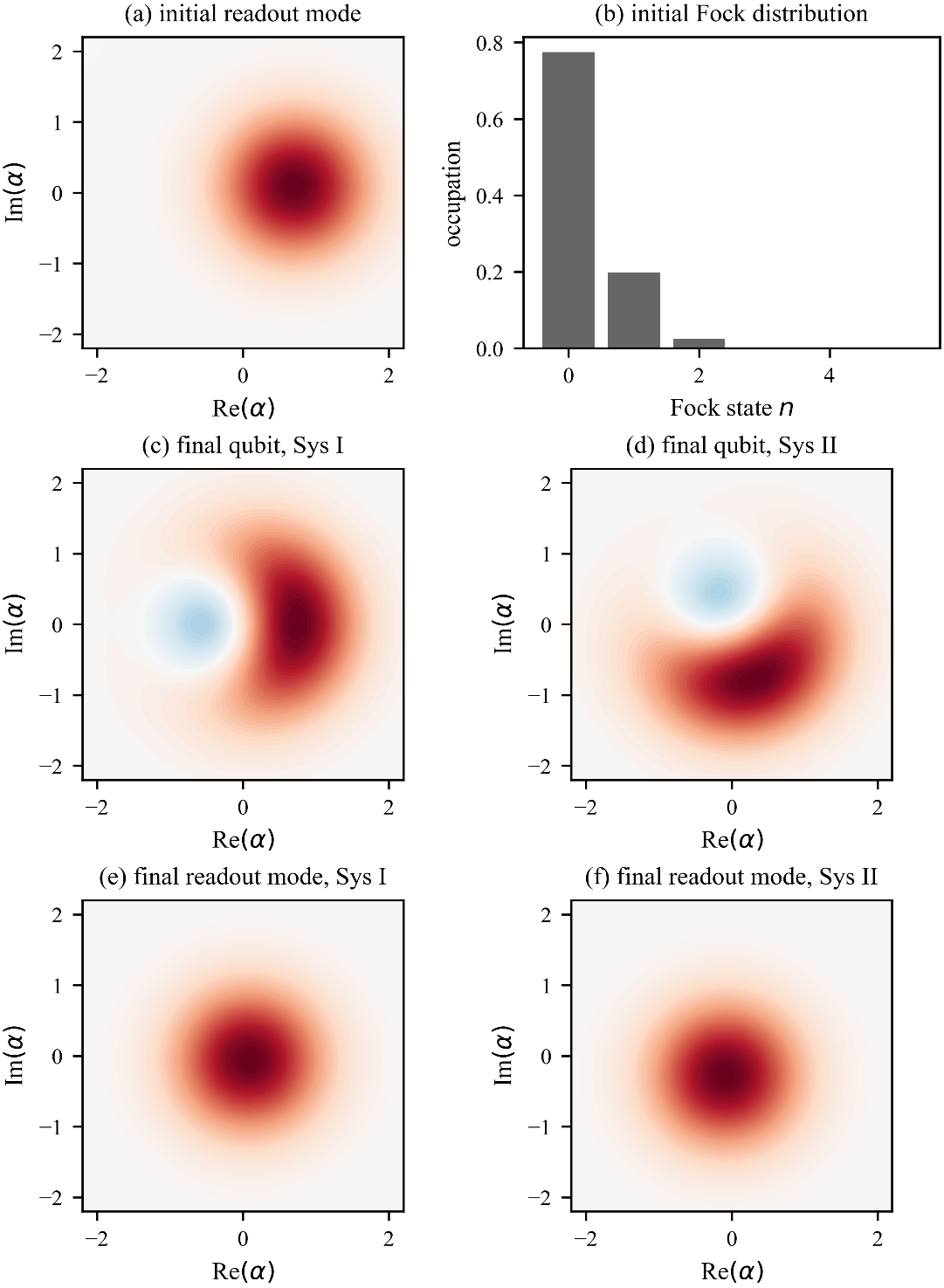}
\caption{Initial and final states of the qubit and of the readout mode, from the master equation, Eq.~(\ref{eq:master}). (a) Wigner function of the initial coherent state of the readout mode, $\alpha_0 = 0.5 + 0.07j$, and (b) its Fock-state occupation. (c), (d) Final qubit states for Sys~I and Sys~II; the transmon is treated as a three-level Duffing oscillator initialised in $(|0\rangle + |1\rangle)/\sqrt{2}$, and the Wigner negativity visible in (c) and (d) belongs to the qubit superposition, not to the readout mode, whose reduced state in (e) and (f) is a mixture of coherent states with a non-negative Wigner function. (e), (f) Corresponding final states of the readout mode for Sys~I and Sys~II. Parameters as in Table~\ref{tab:params}, with NH = 6 per resonator mode.}
\label{fig:A3}
\end{figure}

\begin{figure}[tbp]
\centering
\includegraphics[width=0.57\linewidth]{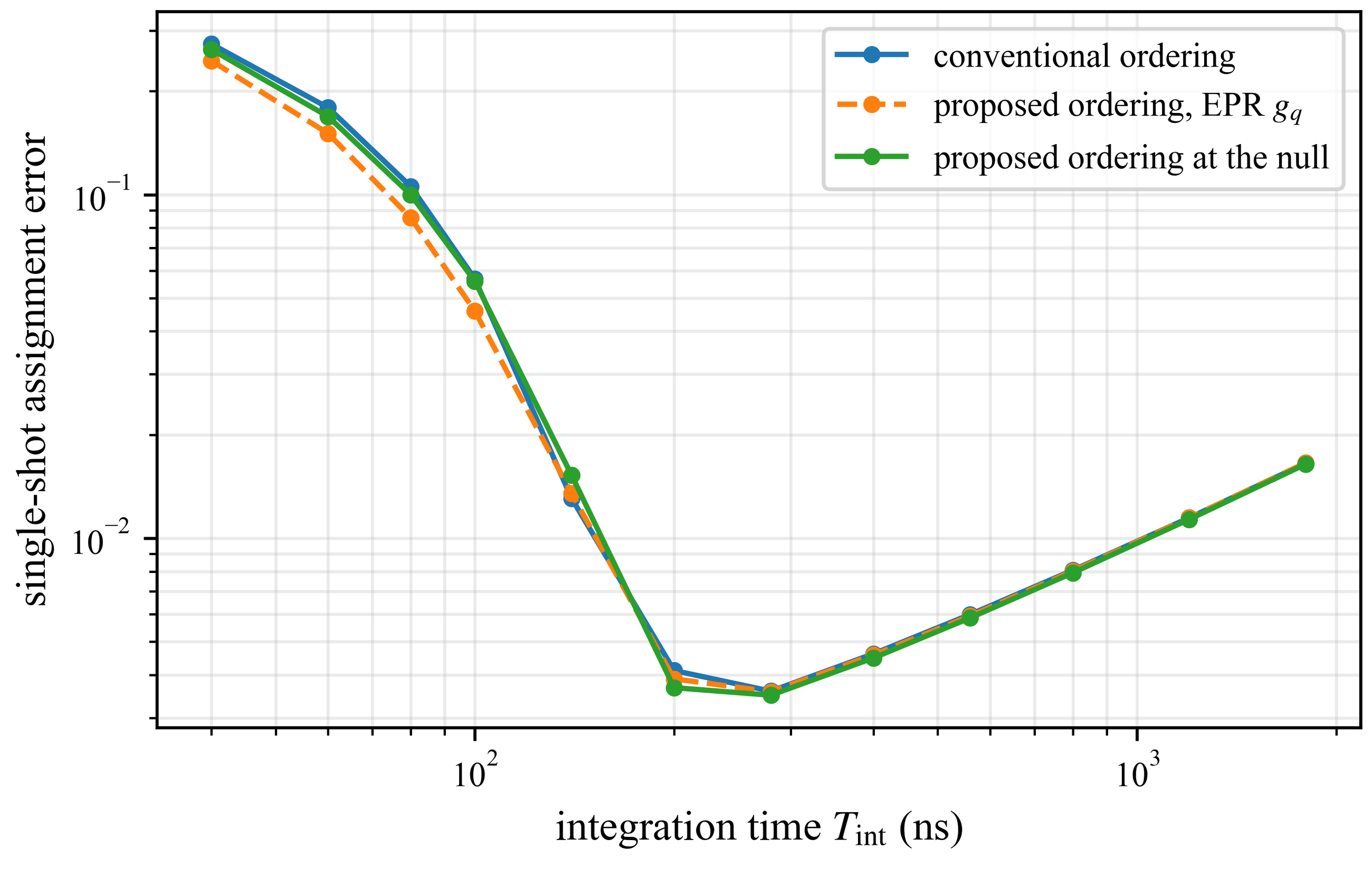}
\caption{Single-shot assignment error as a function of integration time for circuits A, B and C of Table~\ref{tab:circuits}, with $T_1$ from the exact eigenvalue as in Table~\ref{tab:circuits}, with the coupling $g_k$ and the port linewidth $\kappa_r$ re-optimised at every integration time and all other constraints held identical. The rise at short times is the separation error and the rise at long times is relaxation during the measurement.}
\label{fig:A4}
\end{figure}

\begin{figure}[tbp]
\centering
\includegraphics[width=0.72\linewidth]{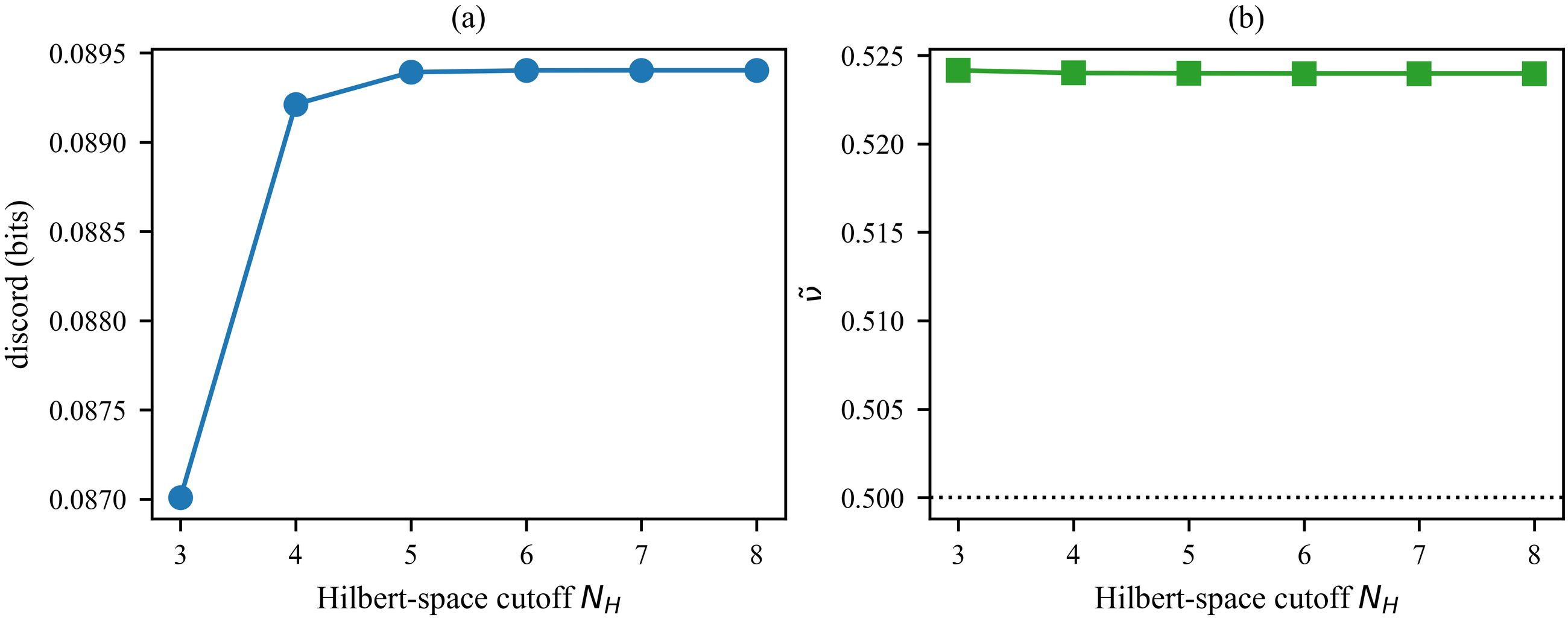}
\caption{Convergence with the per-mode Hilbert-space cutoff NH of (a) the steady-state Gaussian discord and (b) the smallest partially transposed symplectic eigenvalue, at the design point of Table~\ref{tab:params}. Both are converged by NH = 5.}
\label{fig:A5}
\end{figure}

\end{document}